# The Web economy: goods, users, models and policies

Michalis Vafopoulos

05/12/2011

## Abstract

Web emerged as an antidote to the rapidly increasing quantity of accumulated knowledge and become successful because it facilitates massive participation and communication with minimum costs. Today, its enormous impact, scale and dynamism in time and space make very difficult (and sometimes impossible) to measure and anticipate the effects in human society. In addition to that, we demand from the Web to be fast, secure, reliable, all-inclusive and trustworthy in any transaction.

The scope of the present article is to review a part of the Web economy literature that will help us to identify its major participants and their functions. The goal is to understand how the Web economy differs from the traditional setting and what implications have these differences. Secondarily, we attempt to establish a minimal common understanding about the incentives and properties of the Web economy. In this direction the concept of Web Goods and a new classification of Web Users are introduced and analyzed

This article, is not, by any means, a thorough review of the economic literature related to the Web. We focus only on its relevant part that models the Web as a standalone economic artifact with native functionality and processes.

**WSSC:** "*webscience.org/2010/E.1 Economics and Business*"

## Table of Contents

# 1. Introduction

## 1.1. The Web in short

The Web emerged as an antidote to the rapidly increasing quantity of accumulated knowledge in the 20[th] century, which has been caused mainly by scientific progress and digitization technology. Human memory and processing power are extended through the storage and interconnection of online content.

The Web shortened the time, which is necessary for an innovation to become mainstream technology. It took 38 years for telephone technology to reach the threshold of 50 million users, while television needed 13 years, Internet 4 years, iPod 3 years and Facebook just 2 years.

The Web become the new "Promised Land" for quick fortunes and unlimited business growth in the late 1990s because of browsers and search engines that enabled user-friendly navigation. Greed and excessive enthusiasm drove economy in 2001 to a noisy burst of the 5 trillion dollars dot-com bubble.

In mid-00s the Web enabled mass participation and reborn from the ashes of the dot-com bubble. After this hard lesson, the new business models were updated to include advertising revenue from Web navigation and provision of value added services. At this moment, the Web economy is bigger and more robust with new services ranging from search to social networking, virtual entertainment and giant multi-stores. In the demand side, most of the population in the western world is involved in the Web economy. While Silicon Valley is currently focused on the Initial Public Offerings of the leading social networks, President Sarkozy introduces the e-G8 summit and includes the Web in the agenda of the traditional G8 summit.

The Web strengthens the development and democratization processes by empowering people in life-critical functions and enabling participation and transparency.

The Web transformed into the battlefield of a "winner-take-all" fight among titanic firms affecting business and consumer choice in the global economy spectrum. Public and personal infospheres and their interplay are re-invented under new privacy, trust and security laws, ethics and practices. In this new Web ecosystem, researchers and governments are called to create new policy mixtures that will balance market power with personal and social development.

## 1.2. Web and economic research

The emergence of Internet and later the Web, has had an important influence on the research agenda of Economics and Business literature. The massive participation of Users in a variety of social and economic functions created a new terrain of field experiments and analysis concerning consumer behavior, market structure and policy implications. New forms of economic data (e.g. co-purchase networks, real time linked data from Eurostat etc.) enabled researchers to conduct new or existing investigations with less cost. For instance, the estimation of demand for thousands different products is now feasible with only a few weeks of time-series data from Web mass merchants (Chevalier & Goolsbee, 2003).

Yet the available data for research are just a tiny fraction of the collected data from Search Engines, mass merchants, social networks and others in the Web. Contrastingly to physical and life sciences, where massive amounts of open data revolutionized fields like biology and physics, this is not happening for economic and social research (Lazer et al., 2009). The exclusive exploitation of behavioral data in the Web is an issue of primer importance with scientific, economic and social aspects. First, it limits academic research inside the "walled gardens" of companies and government agencies, excluding open scientific research and dialogue. Second, companies that hold data and afford to analyze them have build comparative advantages against (potential) competitors or simply they are selling them for high profit. Finally, privacy and security risks (e.g. personal data leaks, almost-full profiling practices) create negative externalities in the personal and social level, which are not compensated (Vafopoulos, 2006). It is possible that the exclusive and limited data exploitation will become (if it has not already been) the major source of negative externalities in the online world, a form of "digital pollution" in the sense of environmental catastrophe from heavy industries in the traditional economy.

The economic analysis of the Internet and the Web economy follows (with a small time lag) technological improvements and mass phenomena and includes the study of new products, services, business processes, market structures and macroeconomic issues like taxation theory, labor economics, regulatory economics, public goods and development. The first important issue was related to the optimal pricing of Internet traffic (MacKie-Mason & Hal Varian, 1995). The lack of agreement on access pricing was leading in inefficient allocation of limited resources at that time (i.e., bandwidth) (McKnight & Bailey, 1997). The Web as a universal platform for representing and communicating information in digital form initiated the micro-economic analysis of information, network and digital goods including pricing, bundling, sharing, versioning, switching costs, network externalities and standards, economies of scale and scope and antitrust regulations.

## 1.3. The Web Science perspective

The enormous impact, scale and dynamism of the Web in time and space exceed our abilities to observe and measure its evolution process. The complex interplay of social and technological entities occurring simultaneously in the micro and macro level calls for a huge and systematic research effort in order to understand it, model its stylized facts and engineer its future uses in more prosperous ways. Apart from Economics, Web-related studies can be found in many other disciplines such as Computer and Information science, Mathematics, Social and Law studies, to name few.

The common characteristic of these studies is the lack of focus in the Web as a techno-social and standalone artifact. Usually, they refer to conventional questions and apply existing methodologies in their field. But the Web changes some of the underlying assumptions of the human society. The Web depreciates the cost and the institutional barriers to increase the practical potential to exploit the inputs and outputs of the information economy. Peer production emerges, as the third mode of production, a third mode of governance, and a third mode of property. Thus, it is crucial for the future of the online part of our new life to select the fundamental issues, to set new priorities and to concentrate, organize and expand the efforts of Web study.

The trans-disciplinary field in this direction has been entitled "Web Science". Web science is taking the Web as its primary object of study. It is focused in the significant reciprocal relationship among the social interactions enabled by the Web's design, the scalable and open applications development mandated to support them, and the architectural and data requirements of these large-scale applications (Hendler, Shadbolt, Hall, Berners-Lee, & Weitzner, 2008), (White & Vafopoulos, 2011). One of the envelope questions of Web Science is "what changes need to be incorporated in the Web ecosystem to best serve humanity?"

Practically, every discipline is focusing its research efforts on the most important issues during specific periods of time. Nowadays, economists put their efforts to discover new ways for estimating systemic risk because of the severe financial crisis; biologists try to find new personalized cures to diseases after encoding DNA and so forth. Concerning the Web ecosystem, scholars are facing two major research challenges:

1. to preserve and expand the fundamental right of equal and universal online access to information against restrictive political actions and oligopolistic business practices and
2. to accelerate socio-economic development by facilitating life-critical functions in the developing world and by enabling the publication, interlink and re-use of valuable datasets and services in the developed world.

Related issues to Web Economics and Business are indexed under the Web Society (E) category of the Web Science Subject Categorization System (Croitoru et al., 2011; Vafopoulos, 2011a) (Image 1). The Web Society category includes the following perspectives: Economic and Business analysis, Social Engagement and Social Science, Personal Engagement and Psychology, Philosophy, Law and Politics and Governance.

| E. Web Society |
| --- |
| • E.1 Economics and Business |
| • E.1.1 Economics |
| • E.1.1.1 Goods in the Web<br>• E.1.1.2 The Web economy (See also: E.2.4 Peer production)<br>• E.1.1.3 Economics of security, privacy and trust (See also: E.6.2.2 Privacy, E.6.2.3 Trust, E.6.2.4 Security)<br>• E.1.1.4 Antitrust Issues and Policies in the Web<br>• E.1.1.5 Intellectual property and digital rights management (See also: E.5.2 Digital Rights Management)<br>• E.1.1.6 Web-based economic development (See also: C.2.4 Mobile Web technologies, C.5 Semantic Web/Linked Data)<br>• E.1.1.99 Other in Economics |
| • E.1.2 Business |
| • E.1.2.1 E-commerce<br>• E.1.2.2 Business models in the Web<br>• E.1.2.3 Advertising in the Web, sponsored search (See also: D.2.5 Web as a Complex System, Graphs, Networks, Games)<br>• E.1.2.99 Other in Business |

**Image 1: Web Economics and Business are indexed under the Web Society category of the Web Science Subject Categorization System**

## 1.4. Studying goods, users, models and policies in the Web ecosystem

In the first twenty years of its existence, the Web has proven, to have had a fundamental and transformative impact on all facets of our society. While the Internet has been introduced 20 years earlier, the Web has been its "killer" application with more than 2 billion users worldwide accessing some trillion Web pages. Searching, social networking, video broadcasting, photo sharing, blogging and micro-blogging have become part of everyday life whilst the majority of software and business applications have migrated to the Web.

In the present article the term "Web ecosystem" is used to describe three interconnected parts, namely:

   I.     the Internet infrastructure,
   II.    the Web technologies and online content and
   III.   the Users.

The Web is a software application of the Internet infrastructure. Users navigate, create and edit existing content in the Web, i.e. the Web Goods. Web Goods are networked information goods in digital form and build in the Web technologies.

The scope is specific and limited to review a part of the Web economy literature that will help us to identify its major participants and their functions. The goal is to understand how the Web economy differs form the traditional setting and what implications have these differences. Secondarily, we attempt to establish a minimal common understanding about the incentives and properties of Users and Goods in the Web. This article, is not, by any means, a thorough review of the economic literature related to the Web. We focus only on its relevant part that models the Web as a standalone economic artifact with native functionality and processes.

Nine parts compose this study. After the current introduction, the second part is devoted to understanding the properties of goods in the Web. Online bits are considered to be the "cells", the fundamental particles of the Web ecosystem. We build on the tradition of economic analysis of information, knowledge, digital and network goods to analyze the new type of Web Goods. Web Goods are defined as sequences of binary digits, identified by their assigned URI and hypertext format, and affect the utility of or the payoff to some individual in the economy. Their market value stems from the digital information they are composed from and a specific part of it, the hyperlinks, which link resources and facilitate navigation over a network of Web Goods. In Web 1.0 the dominant resource is documents, in Web 2.0 is Users and their contribution and in Web 3.0 is structured data. Our analysis includes the consideration of Web Goods as commodities, search and experience goods.

The third part refers to the Users of the Web network. At the current Web 2.0 era, Users are the protagonists of the cyberspace because they can easily edit, interconnect, aggregate and comment online content as never before. In this part, a simple and comprehensive categorization of Web Users is provided in order to facilitate the comparative analysis of existing literature in economics of the Web. The distinction of Users is based on the motivations and economic impact of their actions in the Web ecosystem. First, Users are partitioned to *Navigators* and *Editors* of Web Goods. Navigators are consuming information by navigating the Web. Editors are producing Web Goods by creating, updating or deleting online content and links. Editors are categorized to Amateur and Professional based on their production incentives. In contrast to *Amateur Editors* (e.g. Wikipedia editors), *Professional Editors* are profit maximizers and take into account direct financial compensations in producing Web Goods (e.g. a blog with paid advertisements). Editors can be further divided, in the basis of their aggregation capability, to Simple and Aggregators. Aggregators are characterized by their automated mechanisms for selecting and presenting Web Goods and are further elaborated to *Search Engines*, *Platforms* and *Reconstructors*. In the last section of the third part, we identify the core function of the Web economy. Briefly, Navigators explore the Web to acquire utility by consuming Web Goods. This navigation creates traffic streams for Editors. Amateur Editors are concerned to attract traffic for their content, even if they do not actually own it (e.g. personal profile page in Facebook). In contrast, Professional Editors, which own or/and administer

Web Goods can transform some parts of this traffic into income through selling it to third parties or advertising or direct sales of both physical and Web Goods. The resulting income acts as an incentive for Editors to update existing and create new Web Goods, producing the new Web network with novel possibilities for Navigators to maximize their utility. Finally, the aforementioned functions are integrated into a more general framework of four interconnected networks, namely: *Users*, *Topics*, *Queries* and the Web.

The fourth part analyzes the characteristics of production and consumption in the Web. What the Web is primarily contributing in the economy is a new source of increasing returns arising from the provision of more choices with less transaction costs in production and consumption.

More choices in consumption are ranging from larger variety of available goods, to online consumer reviews and ratings. This updated mode of more energetic and connected consumption allows consumers to make more informed decisions and provides them with stronger incentives to take part in the production and exchange of mainly information-based goods. But this provision of more choices in consumption is not always coming without compensation. The leading native business model in the Web is the forced joint consumption of online information and contextual advertisements in massive scale. Attention, as approximated by the logged traffic, is the currency of the Web that incentivizes both Amateur and Professional Editors to update and develop the Web network. Attention has become a primer part of the value chain in the Web economy because it can be more efficiently contextualized. The emergence of energetic and connected consumption blurs the borders between production-consumption and (re-) brings in the fore the concept of prosumption. In the production side, many business operations virtualized, went online and become less hierarchical, niche online markets and services emerged and traditional industries revolutionized. Peer Production communities are based on information sharing mechanisms about inputs and outputs, which create public knowledge repositories to store the community's aggregated preferences and expectations. Peer Production as a new form of decentralized inter-creativity outside the traditional market redefines two economic orthodoxies: diminishing marginal productivity and increasing returns to scale.

In the last section, we discuss how digital and Web technologies drive the de-massification of the media by lowering the access barriers to production, distribution and consumption of online information. Moreover, apart from private and public, Peer production emerges, as the third mode of production, a third mode of governance, and a third mode of property.

The sixth part presents four representative models of the Web economy. Despite the fact that there are many research efforts, which address Web-related issues, these models are analyzing the Web as a stand-alone economic artifact. Their primer object of study focuses on the basic economic functions of Web and their implications to consumer's preferences, firms' choices in the Web and the social welfare. Since the presented studies originate from diverse research communities and different systems of symbols and definitions, we analyze them based on the common understanding for Web Goods, Users and core functions of the Web economy that has been built in the second and third parts.

First, the Stegeman model provides an initial step of understanding the transition from mass to network media. It is concluded that firms could widen total surplus by increasing quality, supplying less advertising and reducing access fees. The welfare results are mostly robust to the presence of small to moderate negative externalities from advertising.

Second, the KKPS model is the next representative attempt to account for the basic economic functions in the Web by specifying the interplay of three out of four main factors (Users-Queries, Topics, Web) of the Web function. The KKPS model focus on understanding how the interaction of Users with Search Engines leads to a power law structure of the Web.

Third, the Katona-Sarvary model extends Stegeman's analysis of content exchange between producers and consumers, to hyperlinks exchange among different producers. It focus on the commercial Web, where advertising is used to increase traffic and revenues, not to inform, nor to signal quality or increase brand loyalty. The analysis of hyperlink incentives provides guidance to marketing managers on how to specialize their business models for the Web. In particular, competition in the commercial Web creates motivation for content producers to specialize in specific Topics. The pattern of out-links is different for both advertising and reference links.

Fourth, the Dellarocas-Katona-Rand model for the first time accounts for the economic implications of free reference hyperlinks placement to content nodes. They found, among other results, that linking can sustain market entry of inefficient players, the main benefit of Aggregators to content producers comes from traffic expansion and the presence of Aggregators incurs social costs that must not be overlooked.

The seventh part describes market regulation and antitrust issues in the Web economy. In particular, we examine the basic antitrust issues raised by the "information gatekeepers" of the Web (i.e. Search Engines) and the "infrastructure gatekeepers" of Internet (i.e. ISPs). With respect to first issue, Pollock argues that the Search Engine market is characterized by two stylized facts: (a) a cost structure, which involves high fixed costs and low marginal costs and (b) pure quality competition for Users that is likely to feature very high levels of concentration and under-provision of quality by a single dominant firm. He demonstrates that since the market mechanism cannot provide socially optimal quality levels, there is space for regulatory engagement. Regulatory policies may involve the funding of basic R&D in Web search, or more drastic measures like the division of SEs into two separate parts: "software" and "service". Regarding the second concern about the "infrastructure gatekeepers", a "non-neutral" Internet access raises the following issues in the economic level: two-sided pricing on the Internet, high possibility for prioritization of information packets, identity-based discrimination in delivering information packets, ISPs can determine which of the firms in an industry sector on the other side of the network will get priority, new and innovative firms with small capitalization will actually be excluded from the prioritization auction, ISPs have huge incentives to favor their own content and applications rather that those of independent firms and this may result in multiple fees charged for a single transmission.

The eighth part raises the issue of Web-based development. We discuss the ICTs' role in relation to social inequality and we highlight the development drivers in the networked information economy. It is argued that the one-dimensional direct connection of ICTs with social inequalities should be now replaced by the more relevant question "what changes need to be incorporated in the Web ecosystem to best serve humanity?" The first step in answering related questions should be the identification of fundamental connections between the Web functions and economic development. The second step in understanding the Web's developmental potential is to consider a minimal framework of relevant policies. The final step in exploring Web-based development is to identify some representative initiatives. In this direction, we describe two representative types of projects concerning Web-based development with different tasks. The Web in Society program was initiated by the Web Foundation to enable content sharing about life-critical functions through mobile phones in developing countries. In developed countries, the primary focus in content sharing is to unleash the economic potential of Open Government Data. The final part discusses issues for further research in the Web economy.

## 2. Goods in the Web

In a few years time, the Web has been transformed to an enormous repository and distribution channel of data and information. Web technologies are making information tangible, editable, uniquely definable and compatible to almost any digital format. These functions changed the traditional production, exchange and consumption processes in an unpredictable manner. For the moment, we are walking in an uncharted ground of research.

In our attempt to understand the basic functions of the Web economy, we start by analyzing the characteristics of online content. Online bits are considered to be the "cells", the fundamental particles of the Web ecosystem.

The discussion initiates by providing working definitions about the classical triptych of Data-Information-Knowledge. Because of the fact that the general definitions do not provide useful insights about the specific characteristics of online content, we turn into analyzing the economic concepts of information, knowledge, digital and network goods.

Information and knowledge are increasingly available in digital form and transferred over networks with almost-zero cost. Social networks and information flows through them are becoming partially observable (e.g. Facebook) creating new forms of production and consumption.

The economic analysis of information goods is mainly based on the "informative" function, not the digital nature of information. Similarly, knowledge goods are better describing and facilitating analysis of human capital as an input in the production function. On other hand, the concept of digital goods is a more focused attempt to capture this new reality, because it refers to the information and knowledge that are relevant to the digital economy, but overlooks the transformative power of networks. Respectively, the notion of network goods includes all the goods (physical and digital) that exhibit network externalities, without taking into consideration the special characteristics of digital goods such as nonrivalry, infinite expansibility, discreteness, aspatiality and recombination.

Web technologies provide the technical platform for representing, interconnecting and exchanging addressable digital information in the Internet network. In this new world, in order to understand the novel life cycle of information that is relevant to the self-powered, collaborative and networked economy, we adopt the concept of "Web Goods". Web Goods are defined as sequences of binary digits, identified by an assigned URI and in hypertext format, that affect the utility of or the payoff to some individual in the economy. Their market value stems from the information they are composed from and a specific part of it, the hyperlinks, which facilitates navigation over a network of Web goods.

Section 2.1 briefly discusses the triptych of data, information and knowledge. The next three Sections are analyzing the basic characteristics of information, knowledge and digital goods, respectively. The analysis of network goods begins with the identification of the core elements in network modeling in Subsection 2.5.1. Subsections 2.6.1 – 2.6.4 demonstrate the origins, the types and the main issues related to network externalities. Network externalities in the Web are presented in Subsection 2.6.5. Web Goods are analyzed in Section 2.7. The related discussion involves the various categorizations, the differences with digital goods and the definition of Web Goods as commodities. The last Section describes the relation between search and experience goods and the Web.

### 2.1. Data, information, knowledge

Last decade, digital and communication technologies, and especially the Web, facilitated the explosion in bits production and consumption. In 2010, Google processed 24 petabytes of data per day that have been created from more than 2 billions Users worldwide(G. Munday & O. Munday, 2010). Web Users have access to some trillion Web pages, spending 700 million minutes per month in Facebook, ordering, only in Amazon, 73 items per second and sending 1.3 exabytes from mobile Web devices (G. Munday & O. Munday, 2010).

The Web has structurally changed the representation, communication and transformation of data to information and knowledge. A growing number of diverse disciplines are investigating the concepts of data, information and knowledge (e.g. Linked data (Bizer, Heath, & Berners-Lee, 2009), Information Society (Lyon, 1988), Knowledge-based economy (Harris, 2001) and development (Cooke, 2006), Knowledge Management (Alavi, 2001)).

Much ink has been spilled in trying to clarify the elusive concept of "information". There are various definitions and debates about the so-called hierarchy of data, information, knowledge and wisdom (DIKW)[1] (Rowley, 2007), but let us limit our analysis in the following conceptualizations.

*Data* is a set of discrete, objective facts about events. In an organizational context, data is more usefully described as structured records of transactions (Davenport & Prusak, 2000). *Information* is inferred from data and is defined as data that are endowed with meaning and purpose (Rowley, 2008). Definitions about *knowledge* refer to information having been processed, organized or structured in some way, or else as being applied or put into action. According to Drucker (Drucker, 1989) *"Knowledge is information that changes something or somebody - either by becoming grounds for actions, or by making an individual (or*

---

*an institution) capable of different or more effective action.*" Information could be better conceived as a flow and knowledge as a stock of accumulated information.

The concepts of data, information and knowledge are tightly connected to the new economy (Rowley, 1998). The goal of this article is not to analyze how the DIKW hierarchy changes in the Web, but to understand the economic incentives and mechanisms of information production and consumption in the Web ecosystem.

### 2.2. Information goods

Information has been an important, and at the same time, a problematic concept for economic theory. It is important because is implicitly or explicitly involved in every economic action and problematic because it is difficult to define and quantify due to this ubiquitous and heterogeneous nature (Hirshleifer, 1973). In Economics, information is been considered both as a discrete entity (commodity or good) and as a state of awareness in primary modeling assumptions (Bates, 1985) and constitutes an active research area, usually under the titles "economics of information" and "information economy". In the classification of Economics (AEA, 2010), "information" appears in the Micro-economics [2] and in the Industrial Organization[3] fields.

For the present analysis, information is considered as an economic good, which (a) can be transferred, (b) has some utility, and (c) is capable of having a value attached to it (Bates, 1985).

Information goods have of course always existed, but for most of their history they have been attached to the medium, usually a non-information physical good like stone, book and newspaper. Initially, an "information good" was generally defined as the good, which main market value emanates from the information it contains. The typical examples of information goods included stories printed in books and news in newspapers. First, telecommunication technologies (e.g. radio, TV) and later digitalization enabled the detachment of information goods from the medium of transfer. This change had tremendous effects in the production, exchange and consumption of information and knowledge that could not be fully captured by the traditional conceptualizations. In 1999, Shapiro and Varian (C. Shapiro & H. Varian, 1999) re-defined information goods to be anything that can be digitized (a book, a movie, a record, a telephone conversation). These "potentially digital" information goods may be copied, shared, resold, or rented in order to provide revenues. When such opportunities for sharing are present, the content producer will generally sell a smaller quantity at a higher price, which may increase or decrease profits. Three circumstances where profits increase may be identified:

- when the transactions cost of sharing is less than the marginal cost of production;
- when content is viewed only a few times and transactions costs of sharing are low; and
- when a sharing market provides a way to segment high-value and low value users.

The distinctive characteristics of information goods that have been identified, mainly in Economics and Information Science literature, can be summarized as follows:

- Typical information good has a *high fixed cost* of production but a *low marginal cost* of reproduction. This cost structure implies increasing returns to scale and therefore non-convexity (Levitan, 1982) (C. Shapiro & H. Varian, 1999), (Oberholzer-Gee & Strumpf, 2007)
- Information is an *experience* good. Consumers need to try it to see whether or not it is useful[4].
- Information can be a public or a private good. Information goods are typically non-rival and sometimes non-excludable (C. Shapiro & H. Varian, 1999).

---

[2] Under the subcategory "D8 Information, Knowledge, and Uncertainty".
[3] Under the subcategories "L15 Information and Product Quality; Standardization and Compatibility" and "L86 Information and Internet Services; Computer Software".
[4] For a more detailed analysis about experience goods in the Web refer to Section 2.8.

Information goods are transferable, durable and non-rival like public goods, but, like private goods, their production is costly. In most of cases, reproduction and distribution are inexpensive (i.e. marginal cost is low).

The major issues concerning information goods refer to *versioning, bundling* and *pricing*. The case of monopolistic power in information goods market has been extensively studied in the economic literature since the analysis is more intuitive than under the hypothesis of perfect competition. Bundling is selling two or more different goods in a package for a single price (e.g. Microsoft Office suite). Bundling could be a profitable pricing strategy for information goods since the marginal cost of adding an extra good to a bundle is negligible (Hal Varian & Farrell, 2004). For instance, Bakos and Brynjolfsson (Bakos & Brynjolfsson, 1999, 2000) demonstrate that when the marginal cost of bundling an additional good is low and the value of each good is identically distributed, selling unrelated information goods for a fixed price can be very profitable in the case of a multiproduct monopolist. In such cases, the monopolist can exactly ascertain consumers' willingness to pay for the bundle and thus, to maximize profits and minimize consumer surplus. Versioning is selling different versions of the same good in different prices (e.g. Microsoft Office for Students). Versioning is a second-degree price discrimination widely used in the information goods industry. The prominent advantage is that versioning allows markets to be served that would otherwise not be served, but comes together with the social cost of quality reduction that is necessary to satisfy the self-selection constraint (Schmalensee, 1981), (Hal Varian & Farrell, 2004). Research on discriminatory pricing for information goods includes discrimination on observable and unobservable characteristics, goldilocks pricing, quality adjustment and other methods of extracting consumer surplus (Hal Varian & Farrell, 2004).

### 2.3. Knowledge goods

Knowledge is mainly studied in Economics as technology and technological progress in the production process. First, Marx with his theories of exploitation and accumulation of capital (Marx, 1867) and later Schumpeter, with his theory about innovative entrepreneurship (Schumpeter, 1911), highlighted technological progress as a basic aspect of capitalism. The classical school of economic thought considered knowledge as an implicit way to increase productivity and to succeed economies of scale in production. In 1956, Solow (Solow, 1956) and Swan (Swan, 1956) initiated a new family of economic models that explicitly account for the connections of technological progress and economic growth. Their *exogenous economic growth model* was based on the assumption that technological progress is exogenous to economic activity and led to the paradoxical conclusion that the rate of growth of income per capita in a long-term balanced economy can be explained only by technological progress. The solution came by the *endogenous theories of economic growth* that incorporate knowledge as an endogenous aspect of production (P. Romer, 1986), (Lucas, 1988), (Rivera-Batiz & P. Romer, 1990). Nowadays, the prevailing view about the role of knowledge in economy is that economic growth could be a combined result of the contribution of productive factors and innovation in economic activity. Numerous epistemological approaches exist for the definition and study of the concept of "knowledge" (e.g. cognitive, social etc.). We limit our analysis to the economic aspects of knowledge creation as a dynamic human activity. This activity enters the production in four dimensions (Lundvall & Johnson, 1994):

1) Know- what (facts),
2) Know-why (scientific knowledge),
3) Know-how (skills) and
4) Know-who (networks).

Know-what refers to information about facts that can be easily represented by symbols. Know-why includes the scientific knowledge, such as the laws on how nature, the human mind and society develop. Contrastingly, to the first two types of knowledge, which are observable and can be accumulated through the access of data and information in paper and Web pages, know-how and know-who are tacit or implicit knowledge, usually called human capital, in the sense that is difficult to codify and transfer among humans and can be acquired mainly by education and experience. Know-how describes the capacity and skills of participating in the economic activity. Know-who refers to the ability of procuring the knowledge that resides in social networks.

But does the suits of information and knowledge goods fit the case of online content?

## 2.4. Digital goods
### 2.4.1. Introduction

Economic analysis of information goods is mainly based on the "informative" function, not the digital nature of information. Similarly, knowledge goods are better describing and facilitating analysis of human capital as an input in the production function. For instance, in the economic analysis of information goods, non-rivalry and infinite expansibility have generally viewed interchangeably because each implies increasing returns and therefore non-convexity (Quah, 2003). In contrast to expansibility that represents a restriction on quantity available to society at zero marginal cost over a specific timespan, nonrivalry refers to a restriction on marginal utility. For digital information goods, infinite expansibility always implies non-rivalry (i.e. everyone can have her own copy), but nonrivalry is possible without infinite expansibility (e.g. a live concert is a once-only event that displays nonrivalry but cannot be reproduced in identical copies). Quah (Quah, 2003) explains that:

*"This distinction has implications for Arrow-Debreu equilibrium. With infinite expansibility, the Arrow-Debreu price equals zero, the marginal cost of reproduction in the digital good. But if the first copy of the digital good uses up resources in its instantiation, then a zero price results in market failure: A socially worthwhile good is left unproduced in equilibrium. By contrast, with nonrivalry but only finite expansibility, Arrow-Debreu prices remain positive and, under appropriate conditions, can produce a socially efficient outcome".*

In addition, excludability is a main issue for the analysis of information goods (C. Shapiro & H. Varian, 1999) but it is not essential to digital information goods since it emerges from external enforcement mechanisms (legal or/and technological) not intrinsic to the good itself (Quah, 2003). Therefore, it is needed a new analytical framework that brings to light and synthesizes digital existence to informative nature of goods in the cyberspace. This framework is provided by the definition of *digital goods* as sequences of 0s and 1s that have economic value. Digital goods are distinguished from other goods by five characteristics, namely: nonrivalry, infinite expansibility, discreteness, aspatiality (or weightlessness or spacelessness), and recombination. Digital goods constitute a more focused and effectual concept because it refers to the information and knowledge that are relevant to the new economy. According to Quah (Quah, 2003) the framework of digital goods:

*"...takes the economics of austere high science, technology, and R&D to apply with equal force to videogames, movies, and pop music, as to biotechnology and computer software. In this framework, some digital goods and some parts of the New Economy have a lot to do with knowledge, skills, and productivity; others, hardly at all."*

### 2.4.2. The fundamental characteristics of digital goods

As mentioned above, digital goods, by their nature, are characterized by the fact that are nonrival, infinitely expansible, discrete, aspatial, and recombinant. In the following lines we briefly discuss these characteristics.

### i. Nonrivalry

Most goods are rival in the sense that consumption by one consumer diminishes the usefulness to any other consumer (e.g. if I drink a bottle of water, I exclude everybody of drinking this particular water). Contrastingly to "traditional" physical goods, digital goods are nonrival, in the sense that many users can consume videos, mathematical theorems and social theories without preventing their use from others. The concept of nonrivalry have been put in a central place of economic analysis first by Rivera-Batiz and Romer (Rivera-Batiz & P. Romer, 1990) in their investigation of endogenous growth models and later by Shapiro and Varian (C. Shapiro & H. Varian, 1999) in the context of information goods analysis.

### ii. Infinite expansibility

Despite the fact that nonrivalry and infinite expansibility share the same roots, back in the letter of Jefferson to McPherson at 1813, are two distinct fundamental characteristics of digital goods, with different impact to human behavior (Quah, 2003). Only in the limit case of perfect nonrivalry is equivalent to infinite expansibility (Pollock, 2005). Infinite expansibility was reemerged by David in the context of modeling economic growth as a function of knowledge dynamics (David, 1992). Pollock (Pollock, 2005) defines infinite expansibility as follows:

*"A good is infinitely expansible if possession of 1 unit of the good is equivalent to possession of arbitrarily many units of the good - i.e. one unit may be expanded infinitely. Note that this implies that the good may be "expanded" both infinitely in extent and infinitely quickly".*

The infinite expansibility and almost zero-cost copying of digital media that have been made possible even by technology illiterate users is the main concern and changing force in the business models of music industries.

### iii. (Initially) discrete or indivisible

Digital goods are (initially) discrete, in the sense that their quantity is measured exclusively by integers, and are only instantiate to quantity one. Alternatively, digital goods are not divisible. As Quah (Quah, 2003) explains:

*"Making a fractional copy rather than a whole one, where the fraction is distant from 1, will destroy that particular instance of the digital good."*

Indivisibility can be further refined by the properties of fragility and robustness. In economic terms, a digital good is robust if a sufficiently small and random reassignment or removal of its bit strings is not affecting the economic value of the good (e.g. MP3 compression). Elseways, the digital good is defined to be fragile (e.g. computer software).

### iv. Aspatial

In the limit, digital goods are aspatial in the sense that they are at once everywhere and nowhere. Aspatiality of digital goods is not identical to the definition of aspatiality in Plato's theory of Forms or theory of Ideas (Ross, 1953), which refers to the absence of spatial dimensions, and thus no orientation in space. Digital goods are real bits located in physical devices (e.g. in the disk drive of our laptop or in a Web server). Moreover, aspatiality of digital goods does not connote that physical space is not an important factor. In economy, geography still matters and especially in the production of knowledge-intensive industries, which synchronous face-to-face interactions and critical mass in human capital are critical inputs (Quah, 2000, 2002). The major implication of aspatiality is that transportation costs in the digital or weightless economy are negligible.

### v. Recombinant

Despite the majority of ordinary public goods, digital goods are *"recombinant, cumulative and emergent-new digital goods that arise from merging antecedents have features absent from the original, parent digital goods."* (Quah, 2003).

The above-described characteristics of digital goods can be proliferated through a compatible distribution channel, which will enable complementarity in production, exchange and consumption. The most efficient channel for digital goods has proven to be the Web network.

### 2.5. Networks

In the classical writing Odyssey of Homer, the blind prophet Tiresias advises Odysseus that: *"It is the*

*journey, not the destination."* (Homer, 1952). This poetic prose could be considered as the first recognition that the connecting links are more important than the nodes themselves. Similar ideas are basic parts of Systems theory (Von Bertalanffy, 1968) with wide application in science (e.g. control theory, operations research, social systems theory, systems biology, systems engineering and systems psychology). The growing importance of networks, as the visual and algebraic representation of systems, has been dramatically accelerated by the emergence of digital and Web technologies. Costless information representation and communication enabled the massive formation and maintenance of human networks in global scale. In this new interconnected and globalized world, links and flows through them, in many cases, are becoming an important source of value. Since networks have been created to focus on the analysis of interconnected co-evolving objects, emerged naturally as one of the leading modeling paradigm in many modern scientific fields. But why networks facilitate the modeling of objects of study into a connected system?

### 2.5.1. The core of network modeling

In the following lines, let us focus in a personal view about the core intuition and function of the network modeling process.

After identifying the main object of our study (e.g. goods, humans), we should define which *common property* of these objects (e.g. consumption, friendship) is the most relevant to our research objective (e.g. identify the most important goods in a specific market or people in a social setting). This common property defines the *links* of the network of objects under investigation. A different common property defines an alternative network definition of the same objects. For instance, instead of friendship relation we may define trust relation as the common property in a social network. Optionally, a specific functional form can be employed to evaluate links. For example, trust can scored from 1 to 5.

Additionally, we can consider another *characteristic property* of the object of study that is more "idiosyncratic" in the sense that is not directly connected or depended to other objects (e.g. price, gender). This characteristic property could be called the "evaluation of node". For example, humans can be scored according to their education level and products based on their reputation. Consequently, the minimum structure of a network is composed by:

(a) the main objects of study, i.e. the nodes and
(b) the most relevant to our research objective common property of these objects, i.e. the links.

The full definition of a network contains in addition to (a) and (b):

(c) the evaluation function of the characteristic property of objects (i.e. is called activation in neural networks, potential in electric networks and so forth) and
(d) the evaluation function of the links, i.e. the weight.

A *graph* is defined to be a set of nodes and links and *network* a set of nodes and links evaluated by random variables. Network theory refers to the statistical analysis of nodes and links and their corresponding variables. A good example of efficient network modeling in the Web is provided by the Katona-Sarvary model analyzed in Section 6.4.

### 2.6. Network goods

Economic networks are considered to be a special category of networks in which their main objects of study or/and their links and evaluation functions are related to economic motives, actions and implications. Network goods are nodes in economic networks characterized by a fundamental property: *interdependent positive utility*. The utility of the existing network increases with an additional user. This means that the law of diminishing returns no longer holds, at least until the network has reached its optimal size. The rest of this section is organized as follows. The first subsection defines network externalities and effects. The origins of network externalities are identified in subsection 2.6.2. The third describes the four types of network effects, while the next discuss important issues related to network goods. The last subsection is devoted to network externalities that exist in the Web ecosystem.

### 2.6.1. Network externalities and effects

*Pure network goods* are defined to be goods that derive their entire value from network externalities. Pure network goods have no value in a network of zero size (e.g. telephony, Internet, the Web) (Economides & Flyer, 1998). *Externalities* in economic theory are defined to be the indirect effects of consumption (i.e. demand side) or production (i.e. supply side) activity, that is, effects on agents other than the originator of such activity, which do not work through the price system. In a private competitive economy, equilibria will not be in general Pareto optimal since they will reflect only private (direct) effects and not social (direct plus indirect) effects of economic activity (Durlauf & Blume, 2008). If this indirect effect (or transaction spillover) is beneficial to the other agents is called a *positive* externality and in the opposite case of a cost is called a *negative* externality. For instance, the addition and interconnection of new information in the Web may result positive externalities if it is educational or joyful, or may cause negative externalities if it is privacy threating or libelous. Furthermore, positive network *effects* characterize a good when more usage of the good by any User increases its value for other Users. These effects are also called *positive consumption or demand side externalities.*

In a significant part of literature, the concepts of network externalities and effects are misleadingly used as synonyms. Network externalities in economic analysis are defined to be the externalities involved in network effects. Usually, in the presence of network effects, a User only takes into account his own utility in his decision to join or not the network. The additional utility her joining provides on all other Users is overlooked in his decision (Rapoport, 2011). As Liebowitz and Margolis (Liebowitz & Margolis, 1994) argue:

*Network effects should not properly be called network externalities unless the participants in the market fail to internalize these effects. After all, it would not be useful to have the term 'externality' mean something different in this literature than it does in the rest of economics. Unfortunately, the term externality has been used somewhat carelessly in this literature. Although the individual consumers of a product are not likely to internalize the effect of their joining a network on other members of a network, the owner of a network may very well internalize such effects. When the owner of a network (or technology) is able to internalize such network effects, they are no longer externalities.*

Briefly, network externalities are determined by four factors: expectations and coordination of consumers, switching costs and compatibility of the network good under consideration with the rest of goods (Katz & Carl Shapiro, 1994).

### 2.6.2. The origins of network externalities

Related reasoning to positive network externalities have been initially presented by Theodore Vail, the president of Bell Telephone in 1908, in order to persuade local telephone companies to adopt a single unified standard (Galambos, 1992). The economic term of network externalities remained largely unexplored for almost half century. In 1950, Leibenstein (Leibenstein, 1950) originated the related term of the bandwagon effect, as *"the extent to which the demand for a commodity is increased due to the fact that others are also consuming the same commodity."*

The development of telephone networks was attracted the interest of a growing number of economists and signaled the emergence of an important part of scientific literature. The first studies referred to long-distance telephony and as Rohlfs (Rohlfs, 1974) observed, *"The utility that a subscriber derives from a communications service increases as others join the system. This is a classic case of external economies in consumption and has fundamental importance for the economic analysis of the communications industry."*

During 1980s the literature focused at economics of standardization in the context of network externalities (David, 1985) and (Farrell & Saloner, 1985). Standardization issues, together with Metcalfe's law stimulated considerable interest in related issues and the analysis of network effects was popularized and extended outside the field of Economics. Metcalfe's Law (Metcalfe, 1995) is a rule of thumb, a heuristic, which specifies that the value of a network increase in proportion to the square of the number of nodes. This simplistic approach was inspired by Sarnoff's Law, which states that the value of the network is proportional to the number of its Users, $n$. Metcalfe's Law raised scientific controversies and motivated numerous variations. Briscoe, Odlyzko, and Tilly suggested a more conservative $n \log n$ asymptotic (Briscoe, Odlyzko, & Tilly, 2006). One of the most prominent extensions was the idea of a "group-forming" network or network with many-to-many connections between Users (Reed, 1999). In such cases, the number of possible sets of more than one member is $2^n - n - 1$. Asymptotically, according to Reed's Law the value of the network is equal to $2^n$. Contrastingly, to one-to-many (e.g. TV) and one-to-one (e.g. telephone) networks, group-forming networks are more realistic representations in the case of social networks in the Web.

One of the most important inefficiencies of these mechanistic rules is the equal contribution of each new node to the total network value. To overcome this problem, Beckstrom introduced a more sophisticated approach, which equals the value of a network on the net value added to each User's transactions conducted through that network, summed over all Users (Beckstrom, 2009).

Various scholars from Economics, Mathematics, Engineering and Social studies are applying and contribute to the network effect literature. It is still an active field of study for of variety of disciplines and industrial sectors, including Web business, telecommunications and product standardization (for a survey see (Shy, 2001)).

### 2.6.3. Types of network effects

Network effects can be elaborated in four main categories: direct, indirect, two-sided and learning network effects. In particular:

(a) Direct network effects are incorporating the direct physical effect of the number of consumers on the value of a good (e.g. mobile phones).

(b) Indirect network effects are secondary results of many consumers using the same system and have been defined as "market mediated effects" such as cases where complementary goods (e.g. mobile phone accessories) are more popular or lower in price as the number of consumers of the underlying good (e.g. mobile phone) increases (Liebowitz & Margolis, 1994).

(c) Two-sided network effects (or multi-sided platforms) are evidenced in cases where increases in usage by one set of consumers increases the value of a complementary good to another distinct set of consumers, and vice versa. In this context, indirect network effects could be considered as a type of one-sided network effects. Examples of two-sided network effects are the hardware-software platforms and the Google's advertising platform (Rochet & Tirole, 2003), (Evans, 2003a), (Evans & Schmalensee, 2007). Actually, most information markets are two-sided markets: they coordinate consumption by advertisers and audiences. Supply coordinates demand on both sides of a two-sided market and sets equilibrium prices. In the Web marketplace, attention is a critical part of the value chain, because it is demanded by advertisers and supplied by consumers. On the other side of the two-sided market, production is demanded by consumers and supplied by advertisers.

(d) Finally, the externalities of the learning networks refer to the consolidation of a specific expert knowledge as the network nodes increase (e.g. Linux and Open Office-type software) (Zodrow, 2003).

The different economic implications of direct and indirect network effects were demonstrated by (Liebowitz & Margolis, 1994):

*Indirect network effects generally are pecuniary in nature and therefore should not be internalized. Pecuniary externalities do not impose deadweight losses if left uninternalized, whereas they do impose (monopoly or monopsony) losses if internalized.*

In the case of the Web, the most profound indirect network externality is the multi-sided platforms. A multi-sided platform provides services to two or more distinct groups of customers who need each other in some way and who rely on the platform to intermediate transactions between them. Multi-sided platforms emerge when there is underlying value from getting multiple sides together but transactions costs are high (e.g. eBay decreased the exchange cost for buyers and sellers). In the Web, multi-sided platforms are primarily performing three interrelated core functions. First, they serve as *matchmakers* to facilitate exchange among users. Second, they build *communities* because this makes it more likely that Users will find a suitable match (e.g. Facebook). Third, they provide *shared resources* and *reduce the cost* of providing services to multiple consumer segments (Evans, 2003b). This practice has resulted an ecosystem that consists of interconnecting multi-sided platform businesses (e.g. Google's advertising platform) with excessive market power.

In Economics, the theory of network externalities and effects has extensive applicability and importance in diverse issues like competition, anti-trust policy and regulation, business strategy, innovation and intellectual property.

### 2.6.4. Important issues related to network goods[5]

*(a) Network externalities could be entry barriers*

From the competitive firm's perspective, a rival's network externality is an important entry barrier. Once a firm has the market lead and provides the greatest network benefit to consumers, rivals must not only overcome the price and technology of the leading firm, but must do so by enough to compensate consumers for the network benefit they lose in switching from the leader to the smaller rival. For instance, this is the case for Google as the leading firm in the Search Engine market.

*(b) Network externalities are not identical to lock-in effects*

Lock-in effects emerge when consumers make the rational choice to stick with an existing good rather than to switch to a competitor. Network externalities can create lock-in by making the alternative provider less attractive in terms of the benefit it will provide to the consumer. Even if offered payment or other benefits to switch, a consumer might refuse because of the network benefit she would lose. In such cases, regulatory frameworks should require firms to "share" or to "open" their network externality (through an interconnection or interoperability mandate). But not all lock-in situations arise from network externalities; lock-in can occur even when the competing good is more attractive if the "switching costs" of moving to the better service are too high. Service termination penalties, incompatibility with already-purchased complementary goods and sunk costs are factors that might keep consumers from switching even to better choices. Regulation can limit switching costs by imposing cooperation between rivals (e.g. number portability in mobile telephony operators, technologies and practices for profile and content portability in social networks or other Web applications (Yeung, Liccardi, Lu, Seneviratne, & Berners-Lee, 2009)).

*(c) Network Externalities Pose Regulatory Challenges*

In the case of not sharing the source of network externality, a network industry can be driven toward almost-monopoly. Even in such a case as the market does not end up into monopoly but only one firm delivers the network externality, the question is whether monopoly effects can be weakened without diminishing the network benefit for consumers and without weakening dynamic competition. Network market dominance is not necessarily a result of anticompetitive strategies, but can be arising from consumer preferences. Network monopoly is not, as in conventional markets, clearly bad for consumers because the may receive back benefits from the network externality (e.g. efficient Web search and free services from Google). Thus, regulatory policies have to anticipate the trade-off between conventional price-output objectives and network externalities. A mixing policy approach could be to prevent network externalities and at the same time to achieve the lowest possible price by promoting competition in the market. In the markets of network goods, interconnection of producers and interoperability of goods are primer policy instruments in managing policy objectives. Antitrust measures are necessary where a regulated network operator has an unregulated line of business that is complementary to its regulated service. For example, at the end of 2010, European Commission introduced an antitrust investigation into allegations that Google has abused a dominant position in online search to impose preferential placement of its own services in the advertising market (European Commision, 2010).  In similar cases more intense intervention and strict regulatory framework will be needed to motivate competition in the short run. In regulating network goods there are two types of questions for scholars and decision makers:

1. Should regulators intervene to ensure sustained competition in an evolving network market - either through vertical access (i.e. network neutrality) or horizontal access (i.e. interoperability)?
2. If a network monopoly emerges, should regulators try to limit the duration of the monopoly or otherwise confine its behavior? And if so, when and how?

The normative answer in these questions follow three main policy directions, depending on the type of the market:

1. Divestiture of the monopoly into separate firms.
2. Unbundling or wholesale access to incumbent's facilities (e.g. Internet explorer's case (Economides, 2001)).
3. Licensing of proprietary interfaces to potentially competing platforms.

Each of these involves potentially important economic tradeoffs. In the case of divestiture, if divested entities evolve in ways not fully compatible with each other, economies of scale and network externalities could be vanished. The regulative restrictions of wholesale access, unbundling and licensing of interfaces could deteriorate investment and innovation in the underlying market.

As the Web ecosystem includes and directly or indirectly affects a growing number of markets, consumption interdependencies are becoming complex and the above questions hard to answer. Surely, the World Wide Web Consortium's role as a guarantor that preserves the sharing of network externalities of Web technologies through collaboratively created open standards is fundamental but not enough to discline the rise of almost-monopolies. This strategy is not sufficient anymore to prevent monopolistic practices due to the mass two-sided effects in many markets including mobile Web, video, communication, books, news and mass media industries. Related issues in Web search, media markets and net neutrality are discussed in Sections 7.1, 6.5 and 7.2.3, respectively.

### 2.6.5. Network externalities in the Web

The major value source of the Web is the ability to link resources. In Web 1.0 the dominant resource is documents, in Web 2.0 is Users and their contribution (O'Reilly, 2005) and in Web 3.0 is structured data (Berners-Lee, 2006)(Document_not_found, n d). Contrastingly, to Web 1.0, where the positive network effect arises from the demand side, in Web 2.0 the major source of value stems from the update of online content (i.e. the supply side).

According to Hendler (Hendler, 2009) Web 3.0 extends current Web applications using Semantic Web technologies and graph-based, open data (Figure 1). A potential joint exploitation of all linking spaces in the Web will create enormous social and commercial value (Hendler & Golbeck, 2008).

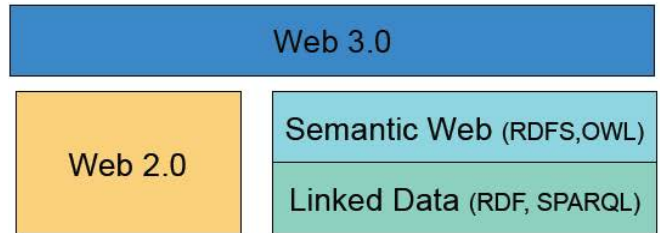

**Figure 1: Web 3.0 extends current Web applications using Semantic Web technologies and graph-based, open data** (Hendler, 2009)**.**

According to Vafopoulos (Vafopoulos, 2011b) Linked Data is an attempt to simplify and spread horizontally throughout the Web the network externalities that exist in Web 3.0. Specifically, two sources of value have been identified for Linked Data technology. First, it enables Web Users to build bidirectional and massively processable interconnections among online data and second, these data are critical enablers for existing infrastructure in the government and business spheres. In particular:

*(a) Building bidirectional and massively processable interconnections among online data*
In contrast to Web APIs, Linked Data mashups are statements that link items in related datasets. As Heath (Heath, 2008) explains: "*Crucially these items are identified by URIs starting "http://", each of which may have been minted in the domain of the data publisher, meaning that whenever anyone looks up one of these URIs they may be channeled back to the original data source. It is this feature that creates the business value in Linked Data compared to conventional Web APIs. Rather than releasing data into the cloud untethered and untraceable, Linked Data allows organisations and individuals to expose their data assets in a way that is easily consumed by others, whilst retaining indicators of provenance and a means to capitalise on or otherwise benefit from their commitment to openness.*"

*(b) Linked Data as an enabler for existing infrastructures*
Public and private entities produce and collect tremendous amounts of data as part of their daily operations. At the same time, increasing investments in IT infrastructure and skills are needed in order to maintain and operate these data on complex hardware and software systems. Linked Data enable the creation of better and massive services for use and reuse for many of these data, driving existing infrastructure in its full potential. For government bodies, Linked Data adoption is focused on open, transparent, collaborative and more efficient governance. For enterprises, the core issue is about effective knowledge management and the implementation of new business models that enable more energetic involvement and collaboration between producers and consumers. There is also significant economic potential in Open Government Linked Data (Pollock, 2009a), which can be used by business as an input to improve existing and to create added value services. This potential can be realized to useful business projects if a certain threshold of data quantity and quality and relevant knowledge is reached. Today, it seems that we are about to approach the triggering point of a virtuous cycle for better services and more involved consumers in the Web economy.

## 2.7. Web Goods
In this part, the concept of Web Goods is discussed. The importance of this new type of goods that reside in the Web is presented in the first subsection. Subsection 2.7.2 refers to the categorization of Web Goods, while the next subsection describes the differences between Web and digital goods. The last section considers Web Goods as commodities.

### 2.7.1. What is a Web Good and why is important
The general classification of Data, Information and Knowledge fails to capture the salient features of the information life cycle in a highly connected digital world. Information can be now digitized (if not digital already) and transferred over networks with minimum cost. Data are transformed to information and knowledge in new ways at global scale. Human networks, and knowledge flows through them, are becoming partially observable (e.g. social networking, institutional Web sites etc.) creating new forms of production and consumption. The concept of digital goods is a more focused attempt to capture this new reality, because it refers to the information and knowledge that are relevant to the digital economy, but overlooks the transformation in fundamental characteristics of information through networks (Vafopoulos, 2011c). Respectively, the notion of network goods includes all the goods (physical and digital) that exhibit network externalities, without taking into consideration the special characteristics of digital goods, namely non-rivalry, infinite expansibility, discreteness, aspatiality and re-combinance. Network goods could be considered as part of the know-how knowledge, as it was defined in Section 2.6. Web technologies provide the technical platform for representing, interconnecting and exchanging addressable digital information in the Internet network (Figure 2). In this new world, we need to understand the novel life cycle of information that is relevant to the self-powered and collaborative and networked economy. To facilitate the understanding and analysis of the Web ecosystem, we adopt the concept of "Web Goods" that is congenital to the online and inter-connected nature of goods in the Web (Vafopoulos, 2011c).

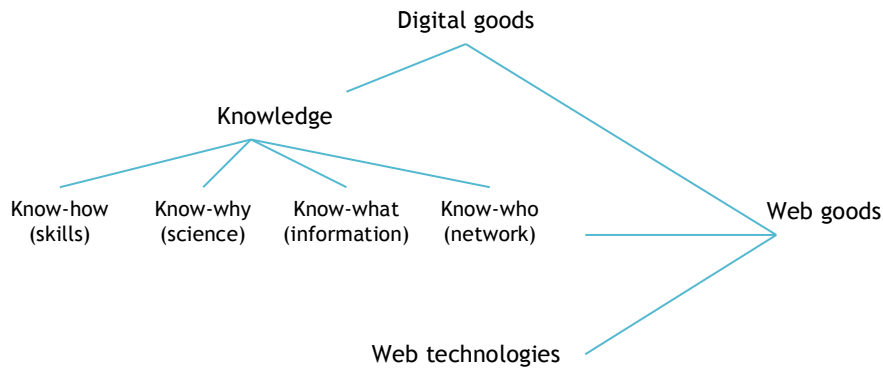

**Figure 2: Web goods are networked digital goods assigned with a URI, represented by Web languages and exchanged through protocols in the Web network.**

*Web Goods (WGs) are defined as sequences of binary digits, identified by their assigned URI and hypertext format, and affect the utility of or the payoff to some individual in the economy. Their market value stems from the digital information they are composed from and a specific part of it, the hyperlinks, which link resources and facilitate navigation over a network of Web Goods.*

For the present analysis, our analytical methodology follows (Quah, 2003) and is based on the background perspective of markets in perfectly competitive Arrow-Debreu equilibrium. Furthermore, analyzes the diacritical characteristics about the Web ecosystem in general or WGs in particular that could affect economic motive and process, rather than studying the Web economy by beginning from ad hoc implicit economic frictions that the Web Economy can then purport to overcome.

Any Web good can be investigated as standalone entity, a member of dyad, a triad or community in a network (e.g. co-purchase network (Vafopoulos, Theodoridis, & Kontokostas, 2011)). As a node of network can be connected to each other and evaluated with various measures such as information content, willingness to pay, number of in-links and out-links. Analogously to digital goods (Quah, 2003), innovation in the Web is the instantiation, i.e., the first creation, of a WG.

### 2.7.2. The basic categories of Web Goods

Web goods (WGs) can be further elaborated in the following categories. *Pure* WGs are the primary focus of the Web Science research because they are defined to include goods that are basically exchanged and consumed in the Web and are not tightly connected to an ordinary good or a service (pre-) existing in the physical world. For instance, a blog entry that comments the market of used cars is a pure WG, but a car sales advertisement is not. According to a production incentives-based categorization, WGs are discriminated in *commercial* (e.g. sponsored search results) and *non-commercial* (e.g. Wikipedia entries). In contrast to commercial, non-commercial WGs are produced outside the traditional market mechanisms of price and property and are based on openness, Peer production and qualitative ex post reward schemes. The economic models presented in Section 6 refer primarily commercial WGs.

Based on their excludability, WGs could be divided to *public* (e.g. Linked Open Data) and *private* (e.g. subscription to the online version of a magazine). There are many different definitions for public goods (Ledyard, 1994). In the present article, a WG is considered to be public if it is non-excludable and is provided by the government, a collective or an individual. A freely available micro-blog entry and the Linked Open Data in data.gov are examples of public WGs. The fundamental importance of public WGs provided by governments has been recently recognized (Pollock, 2009a). However, WGs can be made excludable and in such case, become *purely private goods* through the institutional setting of provision. *Private* WGs are excludable due to a *financial fee* (e.g. subscription paid for a Web service) or a *"personal data"* fee or a *"social" or "membership"* fee. The "personal data" fee refers to the WGs that are requiring User's personal data (usually, only an email address is enough) in order to provide access. This personal information is commonly exploited for marketing reasons and can be overcome by Users through multiple email accounts. The *"social"* fee was introduced by Kumar (Kumar, 2009) in the framework of connected goods. A connected good is defined:

*"as a conspicuous contribution made by an individual that is available for experiencing by the individual's peers in a social network setting. The following properties are shared by connected goods: (1) The contributions are conspicuous information goods and observable by the contributor's peers, (2) the contributor of the good incurs a cost for making a contribution (monetary, time, effort etc.), (3) the contributor need not get a direct or immediate consumption benefit from the good, i.e. the contributor either does not experience the product or service, or already has access to consume the good without making it a connected good, (4) the contributor's friends or peers obtain a benefit from experiencing or consuming the good, hence we refer to them as experiencers."*

For instance, a Facebook profile may not be consumed due to the lack of "friendship" permission from its owner. Connected goods differ from club goods because are not market-mediated but through a social network and include the notion of public goods in the case of a fully connected social network (Kumar, 2009). Connected goods are also distinguished from gift-giving in social contexts because in the latter case, the contributor anticipates real costs for providing gifts and her motivation has been considered to be altruism effects, signaling of wealth etc. Opposingly, in the case of WGs, the contributor does not obtain marginal costs per peer for producing the connected good (Kumar, 2009) and she is basically motived by the exciting experience of navigating into social networks.

### 2.7.3. Differences between Web Goods and digital goods

The Web, as a distribution channel, restricts non-rivalry and infinite expansibility of digital goods because of its limited concurrency capacity and costs imposed by the underlying infrastructure and technologies. As (Pollock, 2005) argues:

*"Now all goods, including intellectual works, must be embodied physically and/or transmitted and/or comprehended to be copied and such activities involve delays as well as utilizing rival goods whose cost is non-zero (though perhaps very small). Thus the act of copying has non-zero cost, the good is not infinitely expansible and is therefore not purely nonrival. For example digital data such as a CD or essay will require resources either to be stored or to be transmitted across a network."*

WGs are initially discrete and indivisible like digital goods. Their distinctive characteristic is that facilitate massive recombinance and consumption in micro-chunks. At the current Web 2.0 era, Users can easily edit, interconnect, aggregate and comment text, images and video in the Web. Most of these opportunities can also be engineered in the personal level. In the Web, nonrivalry, infinite expansibility and indivisibility imply increasing returns of scale and socially inefficient Arrow-Debreu equilibria (if existent) but with different implications each. As mentioned earlier, infinite expansibility in the Web is becoming finite due to network costs (e.g. bandwidth limitations, access fees etc.). According to Quah (Quah, 2003):

*"With expansibility finite, nonrivalry alone presents problems for neither Arrow-Debreu pricing nor social efficiency in perfectly competitive markets. Those conclusions can remain even when the degree of finite expansibility grows without bound, i.e., when digital goods approach infinite expansibility. This limiting result breaks down, though, rather than only at discrete time intervals, whereupon with infinite expansibility market failure then again applies."*

Moreover, indivisibility will result an inefficient Arrow-Debreu equilibrium if only exceeds a minimum threshold scale (Quah, 2003).

Apart from massive information aggregation and recombinance, the Web facilitates and extends aspatiality of digital goods. WGs are characterized, not only by the fact that their transportation cost is low, but that they are accessible from anywhere, anytime. Actually, the Web expands aspatiality and atemporality from local level (e.g. personal hard disk) to global level (e.g. downloadable file link). Intuitively, every Web User can consume all available WGs (freely if they are public and for a fee if they are private) anytime from anywhere. We are all "potential" (or "quasi") owners of each WG, in the sense that may not reside in our memory device but can be downloaded almost instantly. In our point of view, this fundamental expansion of property and existence can be better captured by the concept of virtualization (not to be mixed up with virtual goods (Wikipedia, 2011)). According to Lévy (Lévy, 1998):

*"Virtualization is not derealization (the transformation of a reality into a collection of possibles) but a change of identity, a displacement of the center of ontological gravity of the object considered.... The real resembles the possible. The actual, however, in no way resembles the virtual. It responds to it....Rigorously defined, the virtual has few affinities with the false, the illusionary, the imaginary. The virtual is not at all the opposite of the real. It is, on the contrary, a powerful and productive mode of being, a mode that gives free rein to creative processes."*

Because of this new elastic form of property some human efforts related to time and space have been redistributed to anticipate the third aspect of the Web ecosystem: exploitation of the huge number of choices in consumption and production. Some parts of transaction costs (i.e. transportation and communication costs) have been (re-) invested in making WGs more observable, processable and useful (i.e. the Semantic Web).

### 2.7.4. Web Goods as commodities

Nowadays, a significant part of the available information is produced, exchanged and consumed in the Web. The main difference with existing technological platforms is that information representation and communication could be established in one-to-one, one-to-many and many-to-many forms and in various formats, forms (text, voice, video) and contexts. WGs are the cornerstones of the so-called Economy 2.0 and can be simplistically considered as hyperlinked digital information in a well-defined technological framework. Contrastingly, to information's and knowledge's multiple and controversial definitions and approaches in Economics, WGs qualify as commodities, even according to the strict definition of Debreu (Debreu, 1959). Debreu defines a commodity as a good or a service possessing a stable identity that is completely specified physically, temporally and spatially. Unambiguously, URI technology facilitates a stable identity for every WG, which resides physically in a Web server during a specific period of time. Creation, edit, access and deletion of a WG are recorded in a standardized format under the UTC time measurement (W3C, 1995).

## 2.8. Search and experience goods and the Web

Information goods are characterized as experience goods because you can only tell whether you want to consume the information after you have seen it (C. Shapiro & H. Varian, 1999). Historically, Nelson (Nelson, 1970) was the first to observe that investigation of monopoly market power never considered consumers' information about products as determinant factor. Particularly, he remarked that an asymmetry in consumers' information is an important factor, which facilitates the emergence of monopolies. Nelson, based on the definition of "search goods" as goods whose attributes can be discovered prior to purchase (Stigler, 1961), introduced "experience goods" as the opposite of type of goods. Experience goods are those which attributes cannot be discovered prior to purchasing or consuming the good. Later, Nelson (Nelson, 1974) talked in detail about goods with search and experience attributes using clothes as an example of search good and milk as an example of experience good. Nelson concluded that monopoly is more probable for experience than for search goods and companies pay more for advertising experience goods compared to search goods (Nelson, 1974). The quality of an experience good is initially unknown and is determined during the experience of consumption of the good. Generally, experience goods are characterized by low price elasticity, because a low or discounted price signals the perception of low quality. *But how the Web influences the experience and search attributes of information and non-information goods? Are WGs search or experience goods?*

In Stigler's economics of information (Stigler, 1961) the cost of good that consumer pays is defined as the sum of price plus the cost of searching for it. Since, the Web facilitates the search of relevant information, one can expect that the cost of search is reduced and consequently, the total cost of good that consumer anticipates. However, this is not the only benefit. Klein (Klein, 1998) indicates that the ability of the Web to easily offer information about many goods can change also their experience attributes. She talks about the "virtual experience" which allows the consumer to evaluate basic good's characteristics prior to purchase. By doing so, the experience good is turned into a search good. The author identifies three routes in which this transformation is achieved by examining the case of selling software via the Web. The first route provides the information, which the consumer needs to know about the product through the discussions she can have in user forums where she can obtain knowledge from other users that have experienced the program. The second route provides the information to the customer through third-party reviews. The third route refers to the provision of a free downloadable trial version of the software, which offers the experience to the customer prior to purchase (Stigler, 1961). Klein's research has influenced the posterior research not so much in terms of theoretical aspects, but mainly in terms of practical comparisons between search and experience goods on the Web. Biswas (Biswas, 2004) is among the fewest who tests a set of theories concluding that the Web will influence the traditional economics of information research. Huang et al (Huang, Lurie, & Mitra, 2009) find that despite the fact consumers spend the same amount of time online for both types of goods in order to gather information, for experience goods spend more time per Web page but browse less total number of Web pages than for search goods. They also find that reviews from other consumers prove to be more important for experience than for search goods. Park and Lee (Park & Lee, 2009) find that the effect of word of mouth in the Web, and especially the negative one, is higher for experience goods. On the contrary, Yang and Mai (Yang & Mai, 2010) present evidence that online feedback is more possible to influence consumers in favor of a search than of an experience good. Hao (Hao, 2010) concludes that positive reviews have a greater effect on search goods than on experience goods. However, they find no significant difference in terms of the negative reviews. Girard et al (Girard, Korgaonkar, & Silverblatt, 2003) test basic demographic variables and relate them to search and experience products. They find that gender is significantly correlated to both types of goods, because males prefer to buy online search products like books and experience products like cell phones, while females tend to buy online experience products like clothing. Finally, Moon et al (Moon, Chadee, & Tikoo, 2008) confirm that the Web facilitates selling of search goods.

Navigational ability, collaborative filtering (Sarwar, Karypis, Konstan, & Reidl, 2001), experts' reviews and multimedia representations enrich the search attributes of any good in the Web. For *non-pure* WGs, goods that are tightly connected to an ordinary good or a service exist in the physical world, the magnitude of this search effect depends on the properties of the underlying physical good. For instance, as Klein (Klein, 1998) demonstrated, software could be transformed to a search good if it is properly reviewed and demonstrated in the Web. On the contrary, it is much more difficult to affect the experience attributes of a bottle of wine. In the case of *pure* WGs the predominance between search and experience attributes depends on how close these goods are to information goods. For instance, a story or an article in the Web remains an experience good, despite the fact that several mechanisms exist to enrich initial experience. Reviews, previews and browsing through part of the information give a first impression but the basic attributes cannot be discovered prior to consuming the whole good. On the other hand, a Web-based service could be considered as a search good because, commonly, sufficient information exists about it like screenshots, technical specifications, reviews and evaluations. In general, search attributes are becoming more significant for pure WGs if they include advanced functionalities beyond informativeness.

## 3. Users

### 3.1. A classification of Web Users

Discussion about the search and experience characteristics of WGs and the participatory Web brings into the foreplay the Web Users. In the present study, a simple and comprehensive categorization of Web Users is provided in order to facilitate the comparative analysis of existing literature in economics of the Web (Vafopoulos, 2011d). The distinction of Users is based on the motivations and economic impact of their actions in the Web ecosystem.

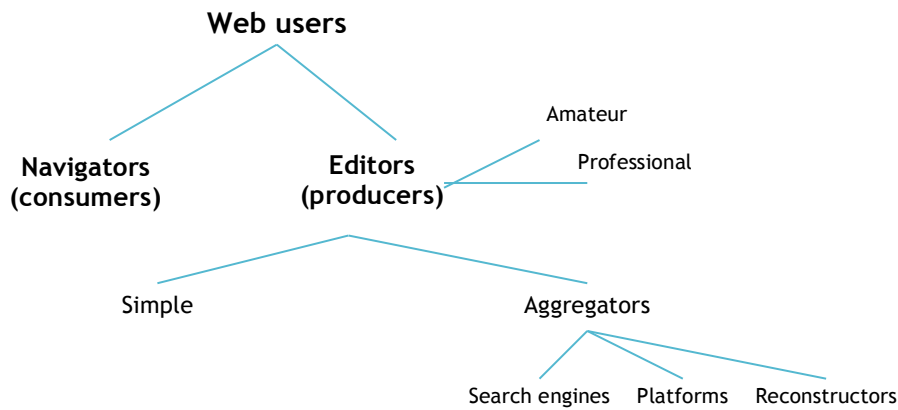

**Figure 3: The Web Users are primarily partitioned to Navigators and Editors of WGs.**

First, Users are partitioned to Navigators and Editors of WGs (Figure 3). Navigators are consuming information by navigating (i.e. browsing, surfing or accessing) the Web network. Editors are producing WGs by creating, updating or deleting online content and links in the Web network. Navigators could be further elaborated in "fact finders", "information gathers" and "browsers" according to information task classification (Kellar, Watters, & Shepherd, 2007), but adds no significant input in the Web's economic assessment. Editors are categorized to Amateur and Professional based on their production incentives. In contrast to Amateur Editors (e.g. Wikipedia editors), Professional Editors are profit maximizers and take into account direct financial compensations in producing WGs (e.g. a blog with paid advertisements). Amateur Editors in not-for-profit community settings (e.g. Open Source Software) are actuated by individual acclaim and reputation-building, which apart from moral reward and self-confidence, may increase their choices to high-paying employment arrangements. This temporal disengagement between effort and reward provide an explanation why Editors may provide knowledge, effort and time for free (Quah, 2003). Amateur Editors in social networking may be motivated by obtaining a higher relative contribution status compared to their peers and future consumption utility from the connected goods provided by their peers (Kumar, 2009). In such cases, Amateur Editors are the initial producers of WGs that are created, bundled and commercialized by a Professional Editor acting as a platform (e.g. Facebook). This massive function in Web 2.0 calls for a function-based distinction among Editors that is economic relevant. Editors can be further divided, in the basis of their aggregation capability, to Simple and Aggregators. Aggregators are characterized by their automated mechanisms for selecting and presenting WGs and are further divided to Search Engines, Platforms and Reconstructors. Their function is more focused on creating content based on filtering and linking existing WGs.

*Simple* Editors are more like Web 1.0 Users, which create content manually and in the case of professionals, monetize their production by advertisement or/and subscription fees. *Search engines* are based on sophisticated algorithms to automatically aggregate, index, classify and (indirectly) commercialize all kinds of existing WGs. *Platforms* are a set of technologies and incentives facilitating Peer production and aggregation under common infrastructure of WGs (e.g., Flickr). Platforms are very important at the Web 2.0 era because they enable Users to collaboratively produce complementary WGs. Commonly, are open-access "walled gardens" in the sense that Users do not pay financial fees to use them, but they produce online content that is difficult or impossible to be transferred to other platforms (lock-in) and their generic code is not open source (Yeung et al., 2009). Most of these Platforms are commercialized by advertisements (e.g. Facebook) and/or subscriptions (e.g. LinkedIn), but also exist not-for-profit platforms that operate as Amateur Editors of the Web network. *Reconstructors* are sophisticated technologies that capacitate the deconstruction, filtering, modification and reconstruction of digital (micro) information into more personalized WGs. For instance, last.fm unbundles music tracks from albums and playlists to reconstruct new playlists based on the collaborative filter matches to User's personal preferences. Reconstructors could be considered as the next generation platforms that are based on semantic processing of WGs (i.e. Semantic Web and Linked Data technologies).

Nowadays, most Platforms and Reconstructors try to consolidate horizontally by adding more functionality (e.g. semantic Wikipedia) and fragment vertically (e.g. LinkedIn professional accounts). The dominant players in Web economy strive to consolidate both horizontally and vertically as Editors of the Web network (e.g. Google News and support of the RDFa specification). Aggregators based their success in the exploitation of the multi-sided platforms by performing three interrelated core functions to reduce the cost of providing services to multiple consumer segments: matchmaking, building communities and providing shared resources[6]. Advertisers in the Web are Professional Editors that create online content to promote consumption of specific goods and services. They can be Simple or Reconstructors. The difference is that Reconstructors interconnect advertising content to the context of a Web page (e.g. AdSense, AdWords).

### 3.2. The core functions of the Web Economy

In today's Web, most of the Users are both navigating and editing online content. How these two fundamental functions interrelate and create economic incentives to result the existing colossal and dynamic network of online information, people and functionalities?

Navigators explore the Web to acquire utility by consuming WGs (Figure 4). This navigation creates traffic streams for Editors. Amateur Editors are concerned to attract traffic for their content, even if they do not actually own it (e.g. personal profile page in Facebook). This function is represented in Figure 4 by the straight line that connects directly traffic to Editors. In contrast, Professional Editors, which own or/and administer WGs can transform some parts of this traffic into income through selling it to third parties or advertising or direct sales of both physical and WGs. The resulting income acts as an incentive for Editors to update existing and create new WGs, producing the new Web network with novel possibilities for Navigators to maximize their utility (Figure 4).

---

[6] For a more detailed analysis of multi-sided platforms refer to Section 2.6.

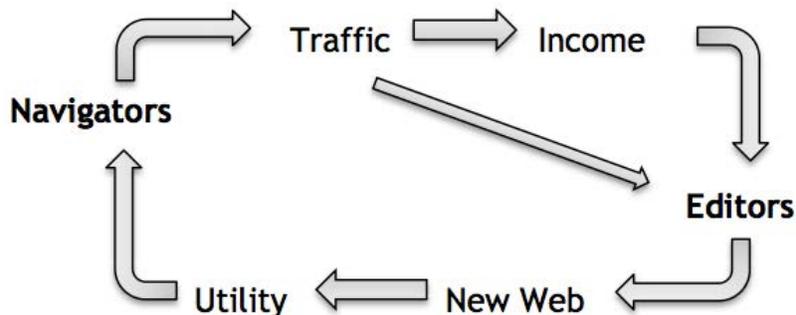

**Figure 4: In the Web economy, Navigators explore the Web to acquire utility. This navigation creates exploitable traffic for Editors, which are motivated to update the existing Web.**

The next step is to include the economic aspect of User's functions into a more general framework of Web functions. This general framework is captured by the contraction of four interconnected networks: Users, Topics, Queries and the Web (Figure 5). A User can access a WG, either by using a Search Engine (Users-Queries-Web) or directly by typing a URI (Users-Web). Users, as have been defined in Section 3, could be modeled to act independently (Stegeman, 2003), (Kouroupas, Koutsoupias, Papadimitriou, & Sideri, 2005a) or to strategically interact (Katona & Sarvary, 2008), (Dellarocas & Katona, 2010). *Query* is the phrasing of a question, usually in terms of a code. The questions are messages expressed as sequences of symbols in the query language. The class of *Queries* has different structures depending upon the interest (for a taxonomic identification of queries you may refer to (Broder, 2002)). A useful description of the class of queries is in terms of graphs or semantic networks. By this way we can include discussions in term of Topics.

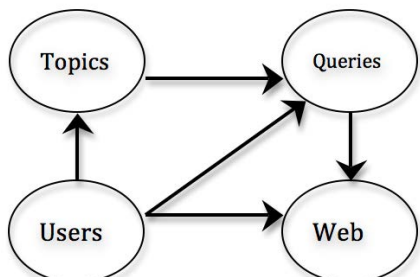

**Figure 5: The Web functions quad graph is defined to be a contraction of four interconnected networks: Users, Topics, Queries and Web.**

Users are explicitly or implicitly interested in specific Topics to navigate and edit the Web. The concept graph is defined with concepts as nodes and semantic relations as links and includes the Topics used in the assessment of WGs. Additionally to the definition provided in Section 3, Search Engines are also considered to be mechanisms that get as inputs Queries and produce results as collections of WGs. The Web graph is formed by WGs as nodes and hyperlinks as links. The Web network accrues from the Web graph in the case of assigning random variables in order to assess nodes and links. For instance, WGs can be assessed by content (e.g. TF-IDF (Yih & Goodman, 2006)). Hyperlink assessment is usually called weight and may be deterministic or random. Modeling and analysis of co-evolution of all four interconnected networks (Users, Topics, Queries and Web) is a difficult task and an important challenge to comprehend and anticipate in a complex network as the Web. There is not yet a model, to the best of our knowledge, which explicitly models all the four interconnected networks and their connections in the Web. An effort in this direction will enable us to design better technologies, experiences and policies in order to exploit the full potential of the Web artifact.

## 4. Consumption and Production in the Web

### 4.1. Introduction

Some economists expected that the Web would gradually lead to perfect information in consumption, acute price competition and pricing at the marginal cost followed by low dispersion (G. Ellison & S. F. Ellison, 2005). The basic arguments were based on lower search and fixed costs, less product differentiation (e.g. location is less important) and "frictionless commerce". There is not strong evidence that many things have changed in these directions in the markets of ordinary goods, since online prices are still dispersed, not much lower than offline (see for instance (Brynjolfsson & Smith, 2000) and (Baye, Morgan, & Scholten, 2006)), and many sectors continue to share oligopolistic characteristics. For some researchers, "obfuscation strategies" (e.g. complicated menus and shipping policies) (Gabaix & Laibson, 2006) and Search Engines' market power prevent potential price reductions. In particular, Search Engines by imposing new types of costs (e.g. Price Per Click) impel retailers to increase their online prices ((Baye & Morgan, 2001). But what actually changed, and not expected at all, was the emergence of new types of consumption and production (e.g. Peer communities), new service sectors (e.g. Software as a Service) and the transformation of existing industries (e.g. mass media). The resulting reconfigurations in the triptych of production-exchange-consumption stemmed from an update in the fundamentals of the economy that the Web brings. Basically, the Web is contributing one major new source of increasing returns in the economy:

***More choices with less transaction costs in production and consumption.***

This source of value arises from the orchestration of digital and network characteristics of goods in the Web. More choices in *consumption* are ranging from larger variety of available goods, to online consumer reviews and ratings. This updated mode of *connected consumption* allows consumers to make more informed decisions and provides them with stronger incentives to take part in the production and exchange of mainly information-based goods. On the other hand, the provision of more choices with less transaction cost in consumption is not always coming without compensation. The leading native business model in the Web is the forced joint consumption of online information and contextual advertisements in massive scale. Section 4.2.1 describes how consumption in the Web economy becomes more energetic and connected. The specific case of co-purchasing networks of a Web mass merchant is presented in Section 4.2.2. Section 4.4 argues how energetic and connected consumption blur the borders between production-consumption and (re-) brought in the fore the idea of prosumption. Moreover, the recent emergence of "social commerce" as a consumer-driven online marketplace of personalized, individual-curated shops that are connected in a network, demonstrates the volatile boundaries among production, exchange and consumption in the Web.

Turning in the *production* side, many business operations virtualized, went online and become less hierarchical, niche online markets and services emerged and traditional industries revolutionized. In the following Sections 5.1.1 and 5.1.2 we present the basic changes in online information production inputs and incentives. Peer Production as a new form of decentralized inter-creativity outside the traditional market is presented in Section 5.1.3. Section 5.1.4 is devoted to the transformation of mass to networked media. *But how these changes are affecting the real economy?*

In my perception, traditional product markets are difficult to change radically and immediately due to their intrinsic characteristics such as existing cost and market structures, institutions, practices and consumer preferences. Contrastingly, the markets where information is involved as primer input and output are transformed with no substantial time lags and are followed by unpredictable outcomes.

### 4.2. Consumption

Due to the rapid penetration of the Web in many technological platforms (e.g. mobile, TV) and social aspects, electronic commerce has become a major activity in ordinary business operations. Almost every firm in the developed world has online presence that describes or/and provides its goods to potential customers. The migration of many business functions in the Web decreased operational costs, primarily, for service-oriented companies.

Electronic commerce is one of the basic components of the Web economy and is gradually becoming an important sector for the entire economy. The Census Bureau of the Department of Commerce announced that the estimate of U.S. retail e-commerce sales for the first quarter of 2011 was $46.0 billion, an increase of 17.5% from the first quarter of 2010 while total retail sales increased 8.6% in the same period. E-commerce sales in the first quarter of 2011 accounted for 4.5% of total sales[7].

The expansion of electronic commerce has attracted many scholars from diverse disciplines such as Economics, Business and Operation Research, Computer and Information science, Law and others (for a review of e-commerce literature see (Ngai & Wat, 2002) (Wang & C. C. Chen, 2010)).

It is beyond the scope of the present article to review and analyze the various dimensions of electronic commerce, but instead we focus on the emergent characteristics of consumption in the Web.

### 4.2.1. More energetic and connected consumption

Trivially, the Web has enabled consumers to access round-the-clock services and to search and compare products, prices, catalogues, descriptions, technical specifications and so forth. Apart from searching and comparing the characteristics of goods and services in the Web, consumers can comment and be informed from others' consumers' purchases and comments. Consumption becomes more connected in the Web. As it was defined in Section 2.6.3 positive network *effects* characterize a good when more usage of the good by any User increases its value for other Users. These effects are also called *positive consumption or demand side externalities*. As consumers become more connected in the Web ecosystem, the network effects are gradually based on the mutual benefits of consumption (Spulber, 2008). *Connected consumption* defines a new form of direct complementarity among consumers. When Navigators consume pure WGs or buy ordinary goods through the Web, reveal and contribute private information about their preferences and expectations, which is beneficial to other consumers if aggregated and made public. These publicly aggregated consumption patterns and comments are valuable in two ways: (a) indirectly, by reducing search and transaction costs (e.g. tags, playlists, collaborative filtering) and (b) directly, by increasing consumption gains (e.g. discovery of complementary goods in co-purchase networks (Vafopoulos, Theodoridis, et al., 2011). Actually, what connected consumers create is not simply content (e.g. product reviews) but *context*. This new contextual framework of consumption emerges through the aggregated personal preferences about WGs in massive scale and facilitates connected consumers to search and navigate the complex Web more effectively, and amplifies incentives for quality. *But how so many and heterogeneous consumers around the globe can coordinate their preferences and expectations?*

In the Coasian world, a small number of consumers can effectively coordinate their preferences through informal agreements and formal contracts to capture the benefits of network effects (Coase, 1960). However, the coordination of large number of consumers requires high transaction costs. Hayek (Hayek, 1991) argued that the price system acts as a coordination device that synchronizes substantial numbers of producers and consumers. Spulber (Spulber, 2008) extended Hayek's analysis of "spontaneous order" to include many other market mechanisms for accomplishing coordination at large. These coordination devices include mass media and marketing, mass communications and observation of other consumers.

In the Web era, Search Engines, social networks and recommendation systems of online retailers (e.g. Amazon, BestBuy) are the most prominent examples of mass coordination devices of consumers' preferences. Search Engines and social networks are general coordination devices that include the full spectrum of preferences. Recommendation systems of the Web merchants are focused in increasing the amount of sales by synchronizing purchasing patterns.

### 4.2.2. Consumer coordination at large in the Web: the Amazon co-purchase network

In the traditional market mechanism, merchants and producers possess all the purchasing history related to their products. Therefore, they exercise the right to exclusively exploit the indirect benefits from

---

consumption externalities. In collaborative Web shopping (or social shopping) this right has been partially shared with consumers. The concept of *social shopping* describes different types of connections among consumers. Practically, four types of social shopping could be identified, namely: group-shopping sites (e.g. Groupon), shopping communities (e.g. Listia), recommendation systems (e.g. Amazon), shopping marketplaces (i.e. online bazaars) and shared shopping (e.g. select2gether). Social shopping needs not to be confused with the emerging term of "social commerce", which refers to connections among sellers.

In social shopping, scholars are mainly investigating the influence of word-of-mouth (see for example (Godes & Mayzlin, 2004)), the technological specifications (Sarwar et al., 2001) and the business implications of recommendation systems. Recommendation systems are basic aspects of the current collaborative Web era since almost every Web commerce business uses information filtering techniques to propose products for purchase like a "virtual" salesperson. Both physical and digital salespersons aim to increase sales, but the difference is that virtual salespersons are only making proposals of the form "most customers that bought this item, also bought" (BLB) based on previous purchases (Vafopoulos, Theodoridis, et al., 2011). Amazon, the biggest Web merchant, is based on a successful item-based collaborative filtering system (Deshpande & Karypis, 2004), providing a wide range of general and personalized recommendations. Specifically, the list of BLB recommended products presents related items that were co-purchased most frequently with the product under consideration. BLB recommendations form the store's co-purchase network and can be represented as a directed graph in which nodes are products and directed links connect each product with its recommended products. In such setup:

*"The virtual aisle location of a product is determined, in part collectively by consumers rather than being chosen based on fees paid by manufacturers, or explicit strategic considerations by the retailer"* (Oestreicher-Singer, 2010).

The idea of examining the Amazon BLB co-purchase network was initiated by Krebs (Krebs, 1999) who proposed the analysis of emergent patterns of connections that surround an individual, or a community of interest, based on book purchases. Oestreicher-Singer and Sundararajan (Dhar, Oestreicher-Singer, Sundararajan, & Umyarov, 2009) extended considerably the analysis by assessing the influence of BLB networks on demand and revenue streams in Web commerce. They also offered new experimental verification about the significance of visible item recommendations on the long tail of commerce in the Web. Recently, in a relevant experimental study (Vafopoulos, Theodoridis, et al., 2011) crawled a set of 226,238 products from all the thirty Amazon's categories, which form 13,351,147 co-purchase connections. They introduce the analysis of local (i.e. dyads and triads) and community structures for each category and the more realistic case of different product categories (market basket analysis). Their main results concerning the purchasing behavior of Web consumers are the following:

- The cross-category analysis revealed that Amazon has evolved into a book-based multi-store with strong cross-category connections.
- Co-purchase links not only manifest complementary consumption, but also switching among competitive products (e.g. the majority of consumers switch from Kaspersky to Norton Internet security suite).
- Top selling products are important in the co-purchase network, acting as hubs, authorities and brokers (or "mediators") in consumer preference patterns.
- Ostensibly competitive products may be consumed as complements because of the existence of compatibility and compatible products that facilitate their joint consumption.

Let us focus on the community analysis of the co-purchase network of products. For a given network, a community (or cluster, or cohesive subgroup) is defined to be a subnetwork whose nodes are tightly connected, i.e. cohesive. Since the structural cohesion of the nodes can be quantified in several different ways, many different formal definitions of community structures have been emerged (Boccaletti, Latora, Moreno, Chavez, & Hwang, 2006). The analysis of community structures offers a deeper understanding for the underlying functions of a network. Figure 6 shows a part of the software products co-purchase network, where different colors indicate different community membership. Different product communities have been identified based on the spin glass community detection algorithm (Reichardt & Bornholdt, 2006). Analysis indicated that the seemingly competitive products of Apple and Microsoft are in reality consumed as if they were complementary. Microsoft (nodes with purple color) and Apple (nodes with orange color) product communities are "mediated" by compatibility like VMware Fusion, Parallels Desktop and compatible products like Office for Mac.

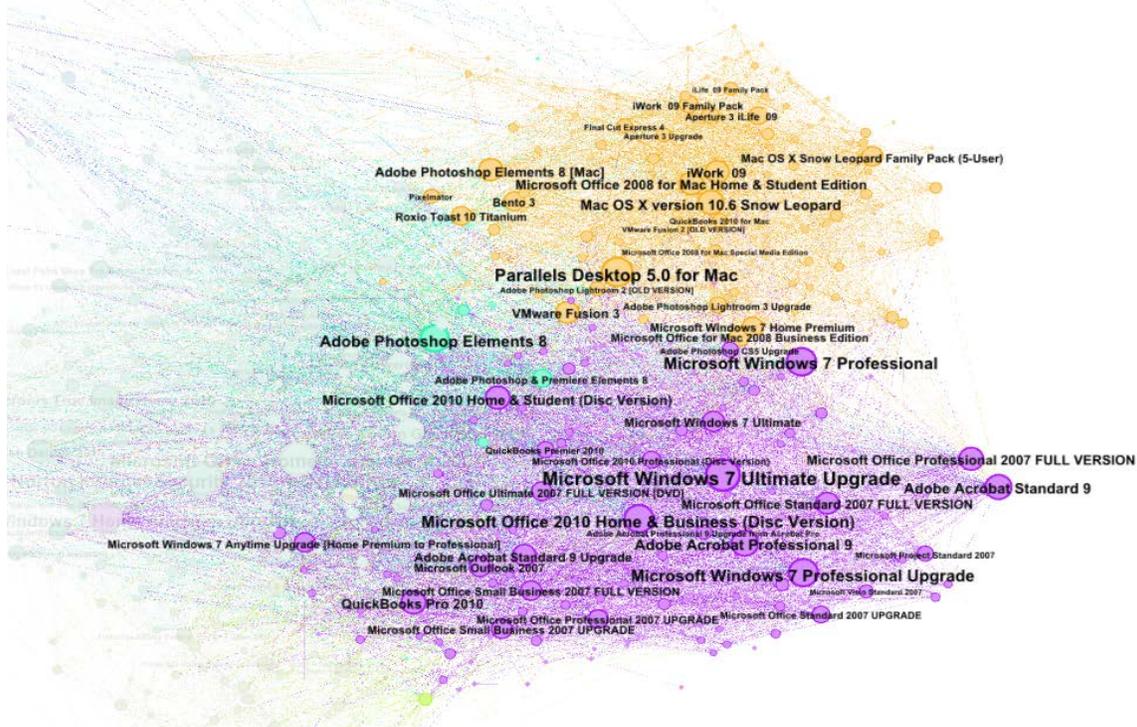

**Figure 6: Microsoft and Apple software programs are consumed as complements, because of compatibility (e.g. Parallel Desktops) and compatible (e.g. MS Office for Mac) products.**

### 4.2.3. Personal data abuse and regulation challenges

As more and more companies are participating in the Web commerce, finding and analyzing consumption patterns is an essential key to their success. Data is the "king" in Web commerce and if are combined with navigational patterns and social networking[8], give to online mass merchants a strong comparative advantage, not only against their direct competitors in the Web but also against to the "brick-and-mortar" retailers. These massive amounts of personal and market data raise concerns about privacy and excessive market power.

The economic analysis of privacy has been initiated before the advent of the Web (Posner, 1978), but become more important with the drastic increase of available data online. Lately, the more encompassing field of the "economics of information security" has emerged to include not only issues like privacy, spam, and phishing, but more general questions such as system dependability and policy (for a comprehensive review see (R. Anderson & Moore, 2006).

To the best of our knowledge, there is no yet scientific investigation in the economic or law literature concerning the excessive market power of Web merchants, which steams from data exploitation. As Clemons and Madhani (Clemons & Madhani, 2010) admit:

---

[8] Since 2010 Amazon users are able to link their Facebook account to their Amazon account. At the outset, this allows Amazon to show you recommendations based on your Facebook interests and activity. http://mashable.com/2010/07/27/amazon-facebook-recommendations/

*"Some digital business models may be so innovative that they overwhelm existing regulatory mechanisms, both legislation and historical jurisprudence, and require extension to or modification of antitrust law."*

An alternative and more generic approach could be to extend the Web architecture to support ex ante information transparency and accountability rather than ex post security and access restrictions (Weitzner et al., 2008):

"*Consumers should not have to agree in advance to complex policies with unpredictable outcomes. Instead, they should be confident that there will be redress if they are harmed by improper use of the information they provide, and otherwise they should not have to think about this at all.*" (Weitzner et al., 2008).

### 4.3. Joint consumption of information and advertisements in massive scale

Spence and Owen first addressed the joint consumption of media content with advertising (Spence & Owen, 1977). Stegeman (Stegeman, 2003) applied their specification to the Web by accounting for constant returns in production, increasing returns in consumption and the forced joint consumption of advertising and online content[9].

Attention has become a primer part of the value chain in the Web economy. Attention, as approximated by the logged traffic, is the *currency* of the Web that incentivizes both Amateur and Professional Editors to update and develop the Web network. *But why attention reserves the central role in the function of the Web economy?*

Attention is massively commercialized in the Web because it can be more efficiently contextualized. The first driver of efficiency is the low transaction cost and the high scalability of information contextualization due to the digital and network nature of Web Goods as they are defined in Section 2.7. The protagonists in the creation of inexpensive and extensively contextualized information are Aggregators (i.e. Search Engines, Platforms and Reconstructors). For instance, advertisements in sponsored search share the same context with algorithmic results. Relatively, purchase recommendations in Web commerce can be related to what the other customers also bought with the product under consideration (e.g. co-purchase network). In collaborative Web sites, opinions, reviews, comments and tags are aggregated under different contexts (e.g. gender, age, interests etc.). As it was mentioned in Section 4.2.1 connected consumers contribute in the emergence of new *context*. This context is created collectively, by markets, networks, and communities and is culturally specific and socially bound.

Information contextualization is indeed an investment in quality that has not been top priority for businesses in the mass media era, since the profit-maximizing strategy was to attract attention and not to invest in production quality.

### 4.4. Moving the borders between production and consumption

Actually, connected consumers, who directly contribute reviews, comments, tags and other online content, are Navigators that become Amateur Editors of the Web. Their content is processed and bundled by Aggregators to create commercial (e.g. Facebook) and non-commercial (e.g. Wikipedia) WGs. In Web 2.0 the traditional triptych of producers-exchange-consumers has been replaced by the prosumption model where consumers contact producers directly or can act, at the same time, as producers (Figure 7). In this setup, Aggregators play the role of exchange by providing bundled WGs for a fee or for free. In this assertion, it is implicitly assumed that the Web is not an exchange itself, but a set of technologies that facilitates the emergence of exchanges and mediating services on top of it[10].

---

[9] The Stegeman model is presented in detail in Section 6.2.
[10] We examine the basic aspects of this mode of production in Section 5.1.3.

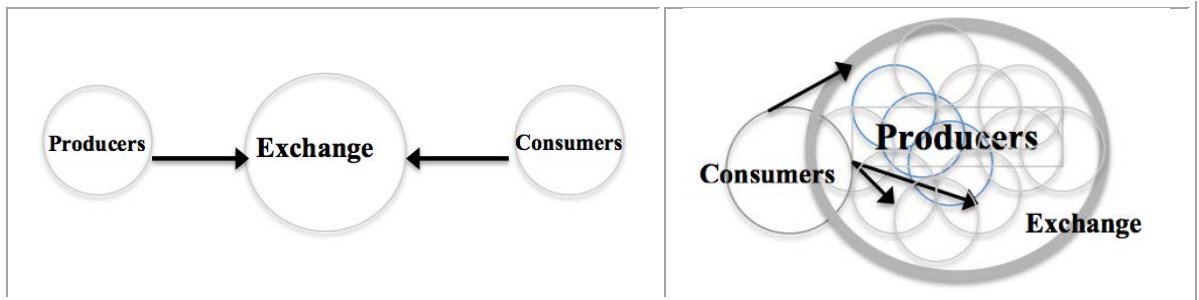

**Figure 7: in Web 2.0 the traditional triptych of producers-exchange-consumers has been replaced by the prosumption model where consumers also contact producers directly or can act, at the same time, as producers**

In the light of these findings, an updated and more general description for online markets is needed. Thus, an online market is considered to be *a set of technologies (e.g. Web platforms, P2P) by which people can exchange decentralized information about preferences and utility (markets) in order to coordinate centralized actions/decisions of consumption and production (firms).*

Amateur Editors, can be also found in the literature as "prosumers" (Ritzer & Jurgenson, 2010) or "produsers" (Bruns, 2008), are participating in the production by crowdsourcing formerly internal business functions (Howe, 2006).

The idea of prosumption initiated by Toffler (Toffler, 1980) arguing that prosumption was important in agricultural societies (first wave) and becomes again a dominant economic model in the information age (third wave). The second wave of industrialization and mass marketization had been interpolated between these two waves by creating a clear production-consumption border. The massification of the participatory Web brought again into the fore the concept of prosumption. Indicatively, Prahalad and Ramaswamy (Prahalad & Ramaswamy, 2004) examined relevant issues on collaboration with customers that co-create value from the marketing point of view and according to Tapscott and Williams (Tapscott & Williams, 2008) Web 2.0 enables crowdsourcing through openness, peering, sharing and acting globally.

As consumption is becoming more energetic and connected, some consumers not only produce tiny parts of information, but also act as exchanges. Lately, is gaining ground the idea of "social commerce". The social commerce marketplaces are characterized by the fact that sellers: (a) are individuals instead of firms, (b) create product collections organized as personalized online shops, (c) interconnect their shops and (d) are paid commissions on sales made by their shops (Stephen & Toubia, 2010). In their investigation Stephen and Toubia (Stephen & Toubia, 2010) concluded for social commerce that *"...(i) allowing sellers to connect generates considerable economic value; (ii) the network's value lies primarily in making shops more accessible to customers browsing the marketplace (the network creates a "virtual shopping mall"); and (iii) the sellers that benefit the most from the network are not necessarily those that are central to the network, but rather those whose accessibility is most enhanced by the network."*

## 5. Production

The declining rents of communication and computation infrastructure enabled billions of people around the globe to participate in the networked information economy. Benkler (Benkler, 2007) argues that:

*"What characterizes the networked information economy is that decentralized individual action— specifically, new and important cooperative and coordinate action carried out through radically distributed, nonmarket mechanisms that do not depend on proprietary strategies—plays a much greater role than it did, or could have, in the industrial information economy."*

WGs overthrow the classical direct, almost one-to-one correspondence of good-consumer (Stegeman, 2003) by providing three different dimensions of output:

- Number of co-producers (supply network externality drive to increasing returns to scale if there are proper mechanisms to enforce and organize incentives, e.g. Peer communities).
- Number of total produced bits (many different WGs bundled under same host).
- Number of concurrent consumers (increasing returns in consumption).

Decentralized Peer production through loosely affiliated self-powered entities is based on a broader baseline of input and output to create a larger range of possibilities for both producers and consumers.

### 5.1.1. Inputs: information and knowledge reloaded

According to Benkler (Benkler, 2007), the production of information is based on three basic inputs: (a) existing information, (b) the mechanical means of conceiving, processing and communicating information and (c) the human communicative capacity. These inputs can be directly associated to the four different types of knowledge[11] proposed by Lundvall and Johnson (Lundvall & Johnson, 1994). Know-what and know-why correspond to existing information and to processing technology, respectively. The tacit form of knowledge (i.e. know-how and know-who) relates to the human communicative capacity. However, Benkler argues that the new Web economy partially changes the mechanisms of creation and accumulation of human capital. Particularly, supports that (Benkler, 2007):

*"...liberation from the constraints of physical capital leaves creative human beings much freer to engage in a wide range of information and cultural production practices than those they could afford to participate in when, in addition to creativity, experience, cultural awareness and time, one needed a few million dollars to engage in information production...The promise of the networked information economy is to bring this rich diversity of social life smack into the middle of our economy and our productive lives."*

### 5.1.2. Incentives: from property to commons

Why people share massive amounts of information in the Web? Why Users abolish their property rights even on some of their personal data?

May be due to altruism, reputation, reciprocity, implicit or explicit social control and money. Or simply because we can do many things better in collaboration than in isolation. Trivially, the institution of property has been recognized as the primer incentive of production and development and a cornerstone in the market economy because it prevents the overexploitation of resources (i.e. tragedy of the commons). Property rights can be further analyzed to four parts:

  i.   The right to use economic resources.
  ii.  The right to modify form and substance of resources.
  iii. The right to benefit from use of resources.
  iv.  The right to transfer resources.

As far as it concerns ordinary goods in the industrial economy, consumers are mainly exercising the first part of the property right. The remaining three parts have been granted (basically through institutions) to producers. Contrastingly, in the collaborative Web, the energetic involvement of consumers in the production of online content was realized through their access to all four dimensions of property.

The production of information-based public goods, such as ideas and knowledge, is primarily based on Procurement and Patronage models to stimulate innovation and societal benefits (David, 1992). Both act as alternatives to property rights, involving the subsidization of creativity by the government or a sponsor in order to avoid the inefficient direct market placement of the product. In practice, funded research schemes disconnect the ex ante incentive of the innovator from the ex post stream of financial flows generated by the innovation (Quah, 2003).

The advent of the Web ecosystem established a new source of production incentives, the so-called Peer

---

[11] A more detailed description is provided in Section 2.3.

Production. This new modality of production is *"...radically decentralized, collaborative, and nonproprietary; based on sharing resources and outputs among widely distributed, loosely connected individuals who cooperate with each other without relying on either market signals or managerial commands. This is what I call "commons-based peer production."* (Benkler, 2007).

In this context, "commons" are considered to be the opposite of "property" in the sense that the underlying resources can be used or disposed of by anyone among some group of Users, under rules ranging from completely free to formal rules that are effectively enforced (Benkler, 2007).

The lack of direct compensation and the temporal disconnection between effort and rewards are the shared characteristics among Peer Production, Procurement and Patronage production models. However, Peer Production is strictly based on collaborative production mechanisms and the emerging supply-side knowledge externalities. Accordingly, Peer Production could be considered as a basic form of production that complements existing and extends David's taxonomy (David, 1992) with the fourth P (Property, Procurement, Patronage and Peer Production). Moreover, apart from private and public, it is argued that Peer production emerges, as the third mode of production, a third mode of governance, and a third mode of property (Bauwens, 2006).

### 5.1.3. Peer Production: decentralized inter-creativity outside the classic market

In today's networked world, the change in User preferences and expectations is tightly related to the rise of Peer Production communities. Peer Production is the creative process of User communities, which collaborate, mainly in the Web, to produce sharable goods. These communities enjoy open access to the means of production, share information about inputs and outputs and create pooled knowledge in order to increase the efficiency of future production. In Peer Production communities private information and preferences are revealed and aggregated without frictions, through explicit (e.g. voting, ranking, pricing) and implicit (e.g. tags, reputation) information sharing mechanisms. Because of the fact that information and preferences are public, transparent choice of inputs and outputs is an efficient coordination rights assignment mechanism. Contrastingly, in traditional business, private hierarchical structures are designed to minimize coordination costs. Peer Production communities could be more efficient than firms or markets if they can operate under less coordination costs in atomizing production. In this context, entrepreneurs have begun to exploit distributed economies of scale in Peer Production on industries with high coordination costs (e.g. social networking, freelancers markets) by providing production platforms.

Peer Production redefines two economic orthodoxies: diminishing marginal productivity and increasing returns to scale. According to the law of diminishing marginal productivity, as resources become less efficient at the margin, productivity of variable inputs declines as quantity increases. The law is valid for "lumpy" goods like cars, computers, and electronics, which are characterized by fragmented divisibility in production tasks, specialized knowledge in multiple activities and high learning, coordination and switching costs of production. Can you imagine how difficult and costly will be to coordinate 10,000 people to produce a car or a computer? Inputs could be indivisible because of cost, technology, regulation, or specialization limitations. But the opposite appears to be the case for a Wikipedia entry or a blog post as many Users make light work at the margin with low coordination costs. In the case of divisible inputs, unlimited number of Users can contribute arbitrarily fine increments of input like a sentence, a paragraph, a photograph or video.

Peer Production communities are based on information sharing mechanisms about inputs and/or outputs, which create public knowledge repositories to store the community's aggregated preferences and expectations. These collective memory mechanisms could be product reviews (e.g. Yahoo! Finance message boards), tutorials and guides, User reputation systems (e.g. eBay seller feedback), collaborative filters (e.g. the Amazon co-purchase network), archived tags and links or various types of community metadata. Knowledge repositories are the key resources and a strong signal for potential productivity gains of joining the community. Users are primarily select Peer Production communities based on their marginal productivity maximization and not on their network size like in network goods. Supply-side knowledge externalities exist if one Editor's adding knowledge to the network increases other Editors' value to it. For Peer Production communities to enjoy supply-side knowledge externalities, knowledge needs to be public, storable, editable, additive and cumulative. WGs within the Peer Production network are public goods produced as an externality of learning-by-doing (Arrow, 1962). Endogenous growth models (Paul Romer, 1994) assume constant supply-side returns at the firm level and increasing returns only at the industry and national level of production. Analysis of Peer Production communities indicates the existence of increasing returns of scale at the micro-production level due to supply-side knowledge externalities. These increasing returns can be sustainable as long as the knowledge pool grows. Increasing returns on Peer Production communities occur if the following virtuous cycle is present: productivity creates new knowledge, which attracts new Users, which increase productivity, which creates new knowledge and so forth. A peer's private productivity is less than his social productivity due to supply-side knowledge externalities. Peer Production happens if Users do not take advantage of other's knowledge sharing (free riding), but contribute to the total productivity of the community. This innovative business model usually fails due to lack of critical mass of Editors and in cases where sharing costs are higher than the cost of atomization.

### 5.1.4. From mass to networked media

During the last decade, digital and Web technologies have been lowered the access barriers to production, distribution and consumption of online information. New property standards (e.g. Creative Commons) and practices are promoting the fair use of content in the networked media. Never before was possible to create, distribute, promote yourself and get feedback for your music, writings or any other online content. Toffler (Toffler, 1980) was one of the first to discuss about the de-massification of the media as a result of information overload and technological advancements. He argues that:

*"A new info-sphere is emerging along-side the new techno-sphere. And this will have a far-reaching impact on the most important sphere of all, the one inside our skulls. For taken together, these changes revolutionize our images of the world and our ability to make sense of it".*

According to the presented analysis in Section 6.5 (Dellarocas & Katona, 2010), the Web competes traditional mass media for Users' attention and the resulting advertising revenues. Actually, in the mass media world, the downstream resources (not the upstream) are relatively scarcer. For mass media attention scarcity is not a basic driver of value creation because barriers to content consumption are high (i.e. limited number of TV channels, radio etc.). The distribution (e.g. broadcasting costs), retail (e.g. spectrum) and production (e.g. infrastructure) scarcities are the major cost centers. Since attention is less expensive to buy through advertisements than costly production, distribution and editing, quality content does not efficiently drives popularity. In the mass media the profit-maximizing strategy is to attract attention and not to invest in production quality. From the market's point of view, mass media firms gain strong first-mover advantages due to the fact that existing high artificial or natural entry barriers limit rivalry. Supply remains limited on both sides of the two-sided market, resulting increases in advertising costs. The dominant strategy is to reuse the same expensive content across many media and markets (e.g. cinema, video, t-shirts, toys etc.). The so-called "Blockbuster Effect" is a strategy to maximize returns on content, based on a more efficient allocation of scarce production resources. In this non-networked world, producers and distributors remain fragmented because production returns do not scale. On the contrary, retailers and marketers capture the most value by consolidating (i.e. acquisitions and partnerships).

## 6. Economic modeling of Web Goods

Issues related to the Web economy have started to occupy a fast growing group of scholars in economic, business, computer science and other disciplines. The main focus of research is online advertising and the underlying mechanisms that transform traffic to revenues. Market structure, firm strategy, market performance and policy implications of the sponsored search market have been started to occupy a growing number of economists.

In the following sections, we have selected to present four representative models of the Web economy. Despite the fact that there are many research efforts, which address Web-related issues, these models are analyzing the Web as a stand-alone economic artifact. Their primer object of study focuses on the core economic functions of Web and their implications to consumer's preferences, firms' choices and the social welfare.

Since the presented studies originate from diverse research communities and different systems of symbols and definitions, we analyze them based on the common understanding for WGs, Users and core functions of the Web economy that has been built in Sections 2.7 and 3.

The first Section provides a brief discussion about advertising in the Web. Section 6.2 analyzes the assumptions and the main results of the Stegeman model. The KKPS model is presented in Section 6.3. The next Section examines the Katona-Sarvary, while the last describes the Dellarocas-Katona-Rand model.

### 6.1. Advertising in the Web

Online advertising is the major source of revenues for Web companies and can be divided into search advertising (or sponsored search), "display advertising" on non-search web pages, classified listings on web sites and e-mail based advertisements. Online advertising has grown rapidly in the last decade[12], but only in 2007 reached the headlines when the U.S. Federal Trade Commission[13] and the European Commission launched in-depth antitrust investigations into Google's acquisition of DoubleClick[14], a service provider to online advertisers. In the micro-economic level, online advertising offers interesting challenges in auction theory, mechanism design and game theory. Per-click pricing (Hoffman & Novak, 2000), envy-free analysis (Hal Varian, 2007), Bayesian equilibrium of the Generalized First Price auctions (Lahaie, 2006), the online allocation problem (Mehta, Saberi, U. Vazirani, & V. Vazirani, 2005) are just a small part of the fast growing literature of algorithmic game theory in the Web (for a collection refer (Nisan, 2007)).

### 6.2. The Stegeman model

Mark Stegeman is one of the first scholars to model the linkage of advertising and information markets in the Web. He extended the existing literature for joint consumption of advertising and other content to account for numerous firms, advertisers and consumers in the Web, with no strategic interactions and demonstrated that firms rely too much on advertising as a source of revenue. This assertion is valid for the mass media and still for a part of the Web where attention (i.e. marketing costs) attracts more funds than quality (i.e. production costs). Stegeman's analysis provides the first step of understanding the transition from mass to network media world. He concluded that firms could widen total surplus by increasing quality, supplying less advertising and reducing access fees. The welfare results are mostly robust to the presence of small to moderate negative externalities from advertising.

Stegeman describes pure commercial WGs as "images", Navigators as consumers and Professional Editors as firms. He also includes a third distinctive part, Advertisers, which are also Professional Editors with different optimizing behavior. There are not strategic interactions between consumers and firms, forming a totally disconnected Users graph. Similarly, WGs are modeled not to be connected with hyperlinks. The processes of production and consumption of WGs create links from Users to the Web, forming the dual Users-Web function graph. He re-introduces the concept of information goods (initially defined by (C. Shapiro & H. Varian, 1999)) as the goods, which

---

[12] UK advertisers now spend £1 in every £4 on Internet advertising http://www.smartinsights.com/digital-marketing-strategy-alerts/uk-online-ad-spend-latest-statistics-released/. In US Digital Ad Spend increased 14% http://www.marketingprofs.com/charts/2011/4291/digital-ad-spend-to-climb-14-social-media-35. In Australia online advertising expenditure grows 17 percent in 12 months http://www.silobreaker.com/online-advertising-expenditure-grows-17-percent-in-12-months-5_2264555522956984333
[13] Statement of FEDERAL TRADE COMMISSION Concerning Google/DoubleClick http://www.ftc.gov/os/caselist/0710170/071220statement.pdf
[14] http://www.nytimes.com/2007/11/13/technology/13iht-webgoogle.8318381.html

*"…resemble nonrival goods in that the quantity produced (e.g., the number of news stories posted) bounds the benefit available to any given consumer but not the number of consumers who can benefit (e.g. the number who access those stories). Like a private good, however, giving access to the marginal consumer (e.g., additional server capacity) may be costly."*

Stegeman's analysis is based on the price-per-impression advertising model, mainly used in broadcasting and publishing industries and adopted by Web 1.0. He is also not accounting for Search Engines and hyperlinks between WGs. By neglecting network externalities, the major value creation mechanism in the Web, concludes that:*"…the elementary properties of information goods make market power almost inevitable."* This assertion contradicts the modern Web market structure, which is dominated by less than ten mammoth global firms with excessive power arising from multiple-platform benefits.

The basic assumptions of the model are presented in Subsection 6.2.1. Editors and Navigators are presented in the next Subsection. Subsection 6.2.3 states the equilibrium conditions and summarizes the main results of the Stegeman model.

### 6.2.1. Assumptions

Stegeman is focused on the role of advertising in correcting the under-provision of information goods. He argues that despite the fact the advertising subsidy can fix underproduction, it is not leading to the optimal quantity of output. His analysis is based on advertising models, which connect advertising to broadcasting (e.g. (Spence & Owen, 1977)) and public goods (e.g. (S. P. Anderson & Coate, 2000)). Most of these models use Steiner's seminal model of program discrete choice assumed a highly active audience in which individuals were persuaded to view only by the presence of a preferred type of content (Steiner, 1952). The basic assumptions of the Stegeman model are the following:

- Firms produce goods under conditions of constant returns to scale and free entry.
- Firms enjoy increasing returns in consumption, because production costs per consumer decline as access increases.
- Firms, advertisers, and consumers are so numerous and small that strategic interactions can be ignored. Demand and supply are differentiable.
- The model is static and continuous. There is no randomness.
- The firms produce goods containing embedded advertising messages. Embedding means that a consumer cannot disentangle the messages from the WG: she must consume them jointly.
- Consumption decision is binary: a consumer chooses only whether to get a WG, not how much she uses it.
- Each unit of output can be consumed by any number of consumers but only once by each consumer.

### Table 1: The basic variables of the Stegeman model

| | |
|---|---|
| **p** | Price of advertising or price of one impression. |
| **q** | Consumption or the quantity of WGs accessed by one consumer. |
| **s** | Access rate or the fraction of consumers who access a given WG. |
| **c** | Quality or the cost of producing one WG. |
| **χ** | Exogenous marginal cost of providing access to one consumer (e.g. cost of running a web server) |
| **f** | Access fee. What one consumer pays to access one WG. |
| **m** | Message density or the number of advertising messages in each WG. |

| x | Output or the total WGs produced. |

### 6.2.2.Editors and Navigators

The major players of the commercial WGs market are the producers (firms), the advertisers and consumers of WGs. Table 1 includes the list of symbols used in the model.

*i.    Firms*

Each firm makes three decisions, selecting:
- its message density $m$,
- its quality $c$ where $c > 0$ is an exogenous lower bound and
- its access fee $f$.

The message density $m$ is the number of advertising messages embedded in each WG that the firm produces. The quality $c$, measured in money units, can be interpreted to include the firm's cost of producing these messages. Firms sell messages to advertisers and provide access to consumers for a fee that may be positive, negative, or zero. The advertiser pays roughly in proportion to the number of consumers who see the message. As Stegeman explains (Stegeman, 2003): *"The combination of constant returns in production, increasing returns in consumption, the advertising subsidy, and the forced joint consumption of advertising and other content raises issues rarely addressed in market theory."*

Each firm aims to maximize its profit, given by the following equation:

$$\pi(c, m, f, s; p) \equiv msp + (f - \chi)s - c \quad \textbf{(1)}$$

The firm's profit function (1) incorporates three parts. First, the advertising revenue is given by the term $msp$ as follows: the firm sells $m$ advertising messages and enjoys $ms$ impressions, at a market price of $p$ per impression. It is assumed that $p$ is determined in an exogenous and competitive market for impressions. The second term $(f - \chi)s$ is the net fee revenue after subtracting the marginal cost of providing access to consumers. The term $c$ is the unit production cost.

*ii.    Advertisers*

Advertisers are small and minimally specified as the exogenous demand for advertising, approximated by impressions. In Stegeman model the content of advertising messages and the nature of their impact is not defined.

*iii.    Consumers*

Consumers access WGs for a positive, negative, or zero fee and jointly consume non-advertising with advertising content. The consumption decision is binary: a consumer chooses only whether to get a WG, not how much to use it. Each unit of output can be consumed by any number of consumers, but only once by each consumer. It is assumed that $A_j \subset [-\frac{1}{2}, \frac{1}{2}]$ represents the measurable set of WGs in the circle that consumer $j$ chooses to consume. Thus, her total consumption is given by:

$$q_j \equiv \int_{A_j} x(t)dt \textbf{(2)}$$

For simplicity and tractability, consumers' behavior obeys the following separable quasi-linear utility function:

$$U_j = \int_{A_j} x(t)[\beta(c(t), m(t)) - \gamma(|t|) - f(t)]dt + \mu(q_j m_j) - \kappa(q_j) \quad \textbf{(3)}$$

where $c(t), m(t), f(t)$ represent firm t's choices of production cost, message density and access fee respectively. $m_j \equiv \int_{A_j} x(t)m(t)dt/q_j$ is the average density of advertising messages embedded in the Web goods in $A_j$, $\beta$: $\mathbb{R}_+ \times \mathbb{R}_+ \rightarrow \mathbb{R}$, $\gamma$: $[0, \frac{1}{2}] \rightarrow \mathbb{R}_+$, $\zeta$: $\mathbb{R} \rightarrow [0, \frac{1}{2}]$ is the inverse of $\gamma$, $\mu$: $\mathbb{R}_+ \rightarrow \mathbb{R}_+$ and $\kappa$: $\mathbb{R}_+ \rightarrow \mathbb{R}_+$ are defined to be exogenous twice-differentiable functions.

$\beta(c(t), m(t)) - \gamma(|t|) - f(t)$ refers to the net benefit that consumer $j$ derives, per WG, from accessing the goods produced by a firm at distance $t$, after deducting the access fee. It is assumed that $\beta_1 > 0$, meaning that consumers prefer higher quality Web goods. There is no specific assumption for $\beta_2$ because consumers may get positive or negative benefits from the consumption of advertising. The function $\gamma$ represents the utility penalty and $\zeta$ its inverse.

The function $\mu$ represents consumer's $j$ benefits from consuming advertising, which are independent of her consumption of WGs and consequently not captured by $\beta$. The argument $q_j m_j$ is the total quantity of messages that she accesses. The function $\kappa$ depicts the opportunity cost of the time spent accessing WGs.

### 6.2.3. Equilibrium

Equilibrium is reached in two stages. In stage one, each firm selects its WG attributes $m$ and $c$ and the access fee $f$. The free entry condition determines total output $x$. In stage two, each consumer, based on firms' decisions in stage one, selects which set of WGs to access. The advertising market clears at price $p$. Since there are no strategic interactions among consumers, no game is played in stage two and the Nash equilibrium takes a simple form. In particular, an outcome (c, m, f, x, q, s, p) is a full equilibrium if it is a consumer equilibrium, satisfies free entry and individual firm optimization. The necessary conditions for equilibrium are the following:

| | | |
|---|---|---|
| $s = 2\zeta(\beta + m\mu' - f - \kappa'(q))$ | **(4)** | [consumer optimization] |
| $p = \alpha(qm)$ | **(5)** | [advertiser optimization] |
| $q = sx$ | **(6)** | [identity] |
| $(pm + f - x)s = c$ | **(7)** | [free entry] |
| $\beta_2 + \mu' + ps\beta_1 = 0$ | **(8)** | [firm FOCs[15] for c and m] |
| $(\beta_2 + \mu')(pm + f - \chi)2\zeta'\left(\gamma\left(\frac{s}{2}\right)\right) + ps = 0$ | **(9)** | [firm FOCs for m] |
| $\beta_2 + \mu' + p = 0$   if f≠ 0 | **(10)** | [firm FOCs for f and m] |
| $\beta_2 + \mu' + p \geq 0$   if f= 0 and WGs are excludable | **(11)** | |
| $\beta_2 + \mu' + p \leq 0$   if f= 0 and WGs are includable | **(12)** | |

where $\alpha$: $\mathbb{R}_+ \rightarrow \mathbb{R}_{++}$ and differentiable with $\alpha' < 0$ and the arguments of $\beta(c,m)$, $\beta_1(c,m)$, $\beta_2(c,m)$, and $\mu(qm)$ are suppressed and understood to take equilibrium values.

WGs are excludable, if positive fees are feasible and includable if negative fees are feasible (e.g. subsidies, coupons). The necessary conditions (4-12) are used only to derive properties of interior equilibria because are not sufficient for a full equilibrium, since the firm's profit function (3) is typically not globally concave in (c, m, f). The expression $\beta_2 + \mu'$ represents the marginal impact of an impression on consumers' utility. For simplicity, if it is assumed that $f \neq 0$ consumers may gain by the presence of advertising messages to the point that $\beta_2 + \mu' < 0$ because holds $p > 0$.

---

[15] First Order Conditions in optimization problems.

The main results of Stegeman model (Stegeman, 2003) are the following:

- In equilibrium, firms set access fees too high at the margin, relative to what would maximize total surplus. This is true even if the equilibrium fee is negative.
- Firms also put too little quality and embed too much advertising into WGs.
- The main exception is that if non-includability holds (no positive fees are feasible), then firms may produce WGs containing too much quality or too little advertising, as they try to compensate for their inability to pay consumers to access their products.
- By collectively reducing the supply of advertising, firms can often increase their own profits as well as total surplus. This is an instance of collusion improving efficiency. Such collusion against advertisers is not always profitable, but it is demonstrated that firms can generally increase profits and surplus by reducing the supply of advertising and simultaneously raising access fees, if such increases are technically feasible. This shows that firms rely too much on advertising as a source of revenue.
- Firms can set non-zero access fees and generally choose a socially inefficient revenue mix: firms could increase total surplus by making a budget-balanced shift away from advertising revenue (which encourages consumption) and toward fee revenue (which discourages consumption). This result stems from the joint production of a WG and the private good of advertising impressions, which creates the possibility of increasing surplus by shifting revenue production from one margin to the other.

### 6.3. The KKPS model

The second attempt to model the basic economic functions in the Web came from Kouroupas et al (KKPS model thereafter) (Kouroupas, Koutsoupias, Papadimitriou, & Sideri, 2005a, 2005b). Kouroupas et al introduced a simple model that begins to capture the main economic issues of the Web. As they indicate: *"...among the most salient, fundamental, differentiating characteristics of the worldwide web (www), unlike past information systems (which were typically confined within the economic interests of a single enterprise) it is created, supported, used, and run by a multitude of selfish, optimizing economic agents with various (and dynamically varying) degrees of competition and interest alignment: Document authors (who want to be clicked and read as widely as possible), end users (who seek the most relevant and helpful information and most gainful opportunities), and search engines (who want to improve their reputation, measured perhaps in terms of user satisfaction)..."*

They first modeled the interplay of three out of four (Users-Topics-Queries-Web) main factors of the Web function by contracting Users and Queries graphs in a single entity and investigated their relationship with Topics and WGs. Their modeling efforts focus on understanding how the interaction of Users with Search Engines lead to a power law Web structure (Barabási & Albert, 1999). The KKPS model calls pure non-commercial WGs as "Documents", since advertising is neglected. Similarly, Users are defined as Navigators, which are posing Queries in the Search Engine and select some of the results. There are no next steps after the first click; there is no real navigation through the Web graph. The Web editing process has also being ignored by the KKPS model. As in Stegeman model there are not strategic interactions between Users, forming a totally disconnected Users graph. Similarly, Topics and WGs are modeled to be totally disconnected graphs. The only connections are between Users and Topics and Topics and WGs.

The basic assumptions of the model are presented in the first subsection 6.3.1. Subsection 6.3.1 states the equilibrium conditions and summarizes the main results of the model.

### 6.3.1. Assumptions

The statistical analysis of the Web graph can be summarized into four major findings: (Bonato, 2005) 1) *on-line property* (the number of nodes and edges changes with time), 2) *power law degree distribution with an exponent bigger than 2*, 3) *small world property* (the diameter is much smaller than the order of the graph) and 4) *many dense bipartite subgraphs*. In the light of these findings (Kouroupas, Koutsoupias, Papadimitriou, & Sideri, 2005a, 2005b) proposed a model of the Web economy, which shows scale-free behavior. Web evolution is modeled as the interaction of WGs, Users and Search Engines. The Users obtain satisfaction (Utility), when presented with some WGs by a Search Engine. The Users choose and endorse WGs with highest Utility and then the Search Engines improve their recommendations taking into account these endorsements, but not the dynamic interdependence of the Utility on the www state. According to the KKPS model, three types of entities are distinguished in the Web, namely: the Documents (i.e. Web pages, created by authors), the Users and the Topics. The corresponding numbers are denoted by *n*, *m* and *k*. For each topic $t \leq k$ there is a WG vector $D_t$ of length *n*, with entries drawn independently from some distribution. The value 0 is very probable so that about *k - 1* of every *k* entries are 0. For clarity (Vafopoulos, Amarantidis, & Antoniou, 2011)(Amarantidis, Antoniou, & Vafopoulos, 2010) denoted by $D_{td}$ the WG components as the relevance of WG *d* for Topic *t* where *t = 1,..., k* is the Topic index and *d=1,..., n* is the WG index. "There are Users that can be thought as simple Queries asked by individuals". "For each topic *t* there is a User vector $R_t$ of length *m*, whose entries also follow the distribution Q with about *m/k* non-zero entries. KKPS simplify the User-Query interplay into the single entity of Users. (Vafopoulos, Amarantidis, et al., 2011) denoted by $R_{ti}$ the components of $R_t$ which represent the relevance of User-Query *i* for Topic *t*, where *t = 1,..., k* is the Topic index and *i = 1,..., m* is the User-Query index.

The WGs are linked with the User-Query set through Search Engines. Each Search Engine indexes Documents according to an algorithm and provides Users-Queries with WG recommendations based on information it has about their preferences for specific Topics. The Search Engine initially has no knowledge of Utility, but acquires such knowledge only by observing User-Query endorsements. In the ideal situation in which the Search Engine knows Utility, it would work with perfect efficiency, recommending to each User-Query the WGs he or she likes the most. The Utility associated with User-Query *i* and WG *d* is described by the matrix *U(i,d)=$U_{id}$*. The value $U_{id}$ represents the satisfaction the User-Query *i* obtains when presented with WG *d*. The simplest linear relation of the Utility, WG and User-Query matrices is the following:

$$U = \sum_{t=1}^{k} R_t^T D_t$$

and as (Vafopoulos, Amarantidis, et al., 2011) showed each element $U_{id}$ of the Utility matrix is:

$$U_{id} = \sum_{t=1}^{k} R_{it} D_{td}$$

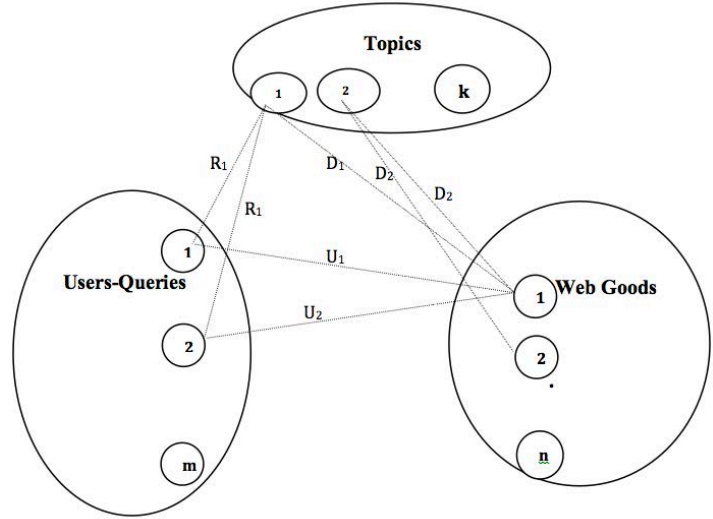

**Figure 8: The Users-Queries, Web goods and Topics tri-graph**

The elements $D_{td}$, $R_{ti}$ and $U_{id}$ define a tri-graph with subgraphs the Users-Queries, the Topics and the Web (Figure 8). In particular, the bipartite subgraph ([$m$], [$n$], $L$) of WGs endorsements by Users-Queries is called by KKPS *the www state*. The Search Engine proposes a number of WGs to the Users-Queries, Users-Queries choose and endorse those that have the highest Utility for them, and then Search Engine makes better recommendations based on these endorsements. It is assumed for simplicity that the number of WGs proposed by the Search Engine is fixed and denoted by $a$, and that the number of endorsements per User-Query is also fixed and denoted by $b$ and that b ≤ α ≤ n.

The basic assumptions of the evolution mechanism in the KKPS model are:

a. every WG is relevant to at least one and even more than one Topic,
b. more than one Users-Queries could be relevant to the same Topic, resulting WGs with in-degree greater than one,
c. a User-Query is concerned with strictly one Topic.
d. no random variable exists

### 6.3.2. Equilibrium

According to (Vafopoulos, Amarantidis, et al., 2011) the structure of the *www state* emerges according to the following 7-step algorithm:

1. A User-Query, for a specific Topic, is entered in the Search Engine.
2. The Search Engine recommends $a$ relevant Documents. The listing order is defined by a rule. In the very first operation of the Search Engine the Documents the rule is random listing according to some probability distribution.
3. Among the $a$ recommended Documents, $b$ are endorsed on the basis of highest Utility. In this way, the bipartite graph $S$= ([$m$], [$n$], $L$) of Document endorsements is formed. Compute the in-degree of the Documents from the endorsements.
4. Repeat Step 1 for another Topic.
5. Repeat Step 2. The rule for Documents listing is the decreasing in-degree for the specific User-Query computed in Step 3.
6. Repeat Step 3.
7. Repeat Steps 4, 5, 6 for a number of iterations necessary for statistical convergence ("that is, until very few changes are observed in the www state" (Kouroupas, Koutsoupias, Papadimitriou, & Sideri, 2005b)).

The main results of the KKPS model are the following:

- It has been experimentally observed that WGs in-degrees are power-law distributed for a wide range of parameters. (Vafopoulos, Amarantidis, et al., 2011) have identified two main mechanisms concerning the origin of the power law distribution of the WGs in-degree: (i) Users-Queries endorse a small fraction of WGs presented $b$, and (ii) assuming a small fraction of multi-topic WGs, the algorithm creates a high number of endorsements for them. The above mechanisms are not exhaustive for the Web graph. Indexing algorithms, crawler's design, WGs structure and evolution should be examined as possible underlying generators for power law distribution.

- The power-law exponent depends on the number $\alpha$ of recommended WGs by the Search Engine, the number $k$ of topics and the number $b$ of endorsed documents per User-Query. (Vafopoulos, Amarantidis, et al., 2011) extended the investigation for different initial random distributions of the in-degree of WGs and for different values of $\alpha$ and $b$ found that the validity of the power law becomes less significant as $b$ increases, both in the case $\alpha=b$ and in the case $b\leq\alpha$, confirming the results of (Kouroupas, Koutsoupias, Papadimitriou, & Sideri, 2005b).

- Concerning the dependence of the efficiency of the search algorithm (price of anarchy (Christodoulou & Koutsoupias, 2005)) on the number $\alpha$ of recommended WGs by the Search Engine, the number $k$ of topics and the number $b$ of endorsed WGs per User-Query they found that the efficiency of the algorithm increases, as $\alpha$ and $k$ increase and $b$ decrease. (Kouroupas, Koutsoupias, Papadimitriou, & Sideri, 2005b) remark that the inverse relation between the efficiency of the search algorithm and $b$ is quite unexpected, since more user endorsements mean more complete information and more effective operation of the search engine. (Vafopoulos, Amarantidis, et al., 2011)extended the investigation confirming the results about $\alpha$ and $k$ and the intuition of (Kouroupas, Koutsoupias, Papadimitriou, & Sideri, 2005b) that increasing $b$ causes the efficiency of the algorithm to increase.

- The "highest endorsement" heuristic, which is a common specialization in the case of both Page rank and HITS, is optimal in the KKPS model.

In the KKPS model, Utility is defined to be time invariant linear function of Users-Queries and WGs, which by construction is not affecting equilibrium when $\alpha=b$. This could be considered as a first approximation, not taking into account the dynamic interdependence of Users preferences. In reality, the evolution of the www state will change both $R$ and $D$. A future extension of KKPS model should account for User behavior by incorporating Web navigating and editing preferences. Additionally, it would be fruitful to offer deeper insight in the Web's business models by incorporating economic aspects in the KKPS model. This could be achieved by introducing valuation mechanisms for Web traffic and link structures and monetizing the search procedure (e.g. sponsored search (Fain & Pedersen, 2006)).

### 6.4. The Katona-Sarvary model

In this direction, Katona and Sarvary (Katona & Sarvary, 2008) created a more realistic model that connects the basic function of selecting links to the scale-free structure of the Web. The Katona-Sarvary model extends Stegeman's analysis of content exchange between producers and consumers, to hyperlinks (in-links and out-links) exchange among different producers. It focus on the commercial Web, where advertising is used to increase traffic and revenues, not to inform, nor to signal quality or increase brand loyalty. The Katona-Sarvary model is based on the marketing literature to specify advertising firms' choices (Dorfman & Steiner, 1954) and supply (Gal-Or & Dukes, 2003) and in Web-related research to model consumers' navigation (Brin & Page, 1998), (Langville & Meyer, 2004) and reference links (Mayzlin & Yoganarasimhan, 2008).

In contrast to Stegeman and KKPS models, Katona and Sarvary account for advertising via links of a network, i.e. advertising effectiveness is endogenous as it depends on the network's structure. Since, the network games literature does not fit for the Web economy (e.g. Bala & Goyal, 2000) consider that individuals in the network are identical and linking to a well-connected person costs the same as connecting to an idle one), they develop a novel approach that results scale-free patterns for in- and out-links in equilibrium.

Users are divided to consumers and producers of online content and Search Engines. Consumers are navigating the Web according to the random-surfer model (Brin & Page, 1998). Producers are Professional Editors of the Web with the twofold role of advertisers and the media. Each Editor is a buyer as well as a seller of advertising. Search Engines are not consider being strategic players but an auxiliary mechanism in finding WGs. The extended Katona-Sarvary model that incorporates Search Engines and Topics analyzes the Web functions as a tri-graph among the contracted Editors-Web graph, the Navigators and the Topics graph.

The analysis of hyperlink incentives provides guidance to marketing managers on how to specialize their business models for the Web. In particular, competition in the commercial Web creates motivation for content producers to specialize in specific Topics. The pattern of out-links is different for both advertising and reference links. WGs tend to purchase advertising links from lower content WGs and the higher content WGs prefer to create more reference out-links.

In the downside, the simplistic random navigation of consumers and the lack of real interaction with content and producers and the absence of total utility analysis, limit the practical applicability of the model. A further investigation of how Users' decisions affect the total utility (i.e. social welfare) could offer useful advice to policy makers on antitrust regulation and development projects. They also, neglect the growing power of Search Engines as major players in the Web economy. Like Stegeman, they misleadingly use the expression "internet business models" for describing the economy of hyperlinks between Web sites.

The basic assumptions of the model are presented in Subsection 6.4.1. The inclusion of Search Engines and Topics in the basic specification of the Katona-Sarvary model is analyzed in Subsection 6.4.2. The last section presents the main results.

### 6.4.1. Assumptions

*(a) Professional Editors: producing in the network*

The Katona-Sarvary model is one of the most comprehensive attempts to model the commercial Web because it is based on a network of online content financed by a market for hyperlinks. Professional Editors and WGs are considered to be identical. Navigators visit these WGs in order to consume their content. Thus, the Users graph is divided into two sub-graphs: Navigators and Professional Editors. Inter-connections between the Navigators and Editors-Web graphs are minimally define how the Web functions. The basic model ignores the existence of the other two parts of the quad-graph, namely Queries and Topics. The Katona-Sarvary model is also extended to include Search Engines and Topics. These extensions model Web functions as a tri-graph among the contracted Editors-Web graph, the Navigators and the Topics graph. As in the rest of the presented models, Search Engines are not considered to be strategic players in the Web ecosystem. It is also assumed that the internal links of every Web site (host graph) are contracted to a single representing node and all the links going out and coming into are assigned to this node. This is a good assumption because each site is owned or/and administered by a single decision maker, the Professional Editor. Professional Editors, which are referred as "sites" or "agents" in the Katona-Sarvary model:

1) are advertisers and the media,
2) are rational economic agents who make simultaneous and deliberate decisions related to the advertising in-links they purchase from each other, based on a specific profit function,

3) are not allowed to strategically choose their out-links. The creation of out-links is only affected by each Editor's pricing strategy, which in turn only depends on the distribution of prices,
4) are heterogeneous with respect to their endowed "content", which may be consider as their inherent value in the eyes of Navigators,
5) generate revenue from two sources:
   a. by selling their content to consumers and
   b. by selling links to other sites,
6) are facing an exogenous Price Per Click that is an increasing function of the originating Professional Editor's content and
7) are producing single-topic content.

The sixth assumption of the initial model is relaxed later by assuming that the Price Per Click (PPC) depends on content. Equilibrium prices are set in a two-stage game, where prices are set first, followed by the purchase of links. The last assumption is updated by the inclusion of Search Engines and a reduced form profit function for Editors with multiple content Topics.

*(b) Consumers: real navigators*
A specific part of Users, the Navigators, browse the Web graph by clicking on hyperlinks. Navigators are considered to be potential consumers of the WG's content (i.e. online information) with some probability $\rho$. The following simplistic assumptions are made for the consumers of WGs:

1) navigate the Web according to a random process, which is nevertheless closely linked to its network structure,
2) have homogenous preferences for WGs and
3) consume all of the WG's content with certainty (probability $\rho = 1$), without loss of generality.

The second assumption will be relaxed later by the extension of heterogeneity in consumers' interest for certain Topics of content. Nevertheless, Editors are not allowed to influence this interest.

**Table 2: The variables employed in the Katona-Sarvary model**

| | |
|---|---|
| $n$ | Total number of WGs |
| $\delta$ | The damping factor of PageRank algorithm |
| $[M]_{ij}$ | Transition probabilities matrix |
| $[U]_{ij}$ | $= \frac{1}{n}$, matrix |
| $q$ | Fixed Price Per Click (PPC) |
| $q_i$ | $= q(c_i)$ PPC of WG $i$, endogenous and increasing function of content |
| $r_i$ | number of visitors at WG $i$ per unit time (also the PageRank of WG $i$) |
| C | cost per-visitor for every WG |
| $r_iC$ | total cost for WG $i$ |
| $p_i$ | total price of an advertising link from WG $i$ |
| $\kappa$ | maintenance cost of reference links |
| $c_i$ | content of the WG $i$<br>i.e. the gain from a consumer or consumers' willingness to pay for this WG |
| $r_ic_i$ | Total income of the WG $i$ |
| **Extended to include Topics** | |
| D | Total number of Topics |
| $\mathbf{c_i}$ | $c_i^1, c_i^2, \ldots c_i^D$ content vector of WG $i$ for Topics 1, 2, …, D |
| $w$ | Proportion of consumers interested in the different Topics or the probability distribution on content dimensions describing the interest of a randomly selected consumer |
| $\mathbf{wc_i}$ | Weighted average content of WG $i$ |
| $r_i\mathbf{wc_i}$ | Total weighted income of the WG $i$ |
| **Extended to include Search Engines** | |
| $b$ | The proportion of consumers which navigate according to the random surfer hypothesis |
| $1-b$ | The proportion of consumers which use a Search Engine in every navigation step |
| s | Search Engine index of WGs |
| $\mathbf{C_i}$ | The content vector truncated by the Search Engine by setting to zero the dimensions that do not make it in the top $s$ ranks |
| $E_i$ | $\equiv b\mathbf{wc_i} + (1-b)\,\mathbf{wC_i}$, The modified average content |
| $d_i^{out}$ | out-degree of WG $i$ $(=d_i^{out_R} + d_i^{out_A})$ |
| $d_i^{in}$ | in-degree of WG $i$ $(=d_i^{in_R} + d_i^{in_A})$ |
| $d_i^{out_R}$ | out-degree for reference links of WG $i$ |
| $d_i^{in_R}$ | in-degree reference links of WG $i$ |
| $d_i^{out_A}$ | out-degree for advertising links of WG $i$ |
| $d_i^{in_A}$ | in-degree advertising links of WG $i$ |

For the basic model, the goal is to determine the number of consumers visiting each WG, in a given unit of time. In this direction, it is adopted the "random surfer" hypothesis of the Google's Page Rank algorithm (Brin & Page, 1998). As in Stegeman model, the total number of consumers is normalized to 1 and is distributed equally between $n$ WGs. All the consumers follow a random navigating behavior in every step. Starting from WG $i$, with probability $\delta$, she randomly follows a link driving out from that WG or remains there, selecting each of these $d_i^{out} + 1$ options with equal probability. Iteratively, she moves to a random WG with probability $1 - \delta$, where $\delta$ is defined to be the "damping factor" (commonly, set to 0.85 or approximately to an expected "surfing distance of seven links). Table 2 includes the list of symbols used in the model. Katona and Sarvary (Katona & Sarvary, 2008) prove that the Page Rank of each WG (i.e. the proportion of visitors reaching it) is the weighted average of two matrices (M and U) each representing a different random process. Formally:

$$r = \delta r M + (1-\delta) r U = r(\delta M + (1-\delta)U) \quad \textbf{(1)}$$

where $[M]_{ij} = \begin{cases} \frac{1}{d_i^{out}+1} \; if \; (i \rightarrow j) \\ 0 \; otherwise \end{cases}$ and $[U]_{ij} = \frac{1}{n}$

M includes the transition probabilities across interlinked WGs, i.e. the navigating patterns of consumers along the links of the network. Hence, it describes the structure of the Web. On the other hand, U contains a process that scatters Navigators randomly around to any of the WGs. These two matrices are weighted by the damping factor, $\delta$, providing a simple but consistent description of how traffic is distributed across WGs for any given link structure of the network. If it is assumed that WG $i$ enjoys traffic $r_i$, then the total number of Navigators visiting the particular out-link is $\frac{\delta r_i}{(d_i^{out}+1)}$ and the total price of an advertising link from WG $i$ is $p_i = \frac{\delta r_i q_i}{(d_i^{out}+1)}$

Traffic, i.e. the number of clicks on a particular hyperlink is given by equation (1). Based on the above assumptions, the profit of the producer of WG $i$, for a given network structure and navigation patterns, is composed by three parts: the net income from its consumers' traffic (revenue $r_i c_i$ minus cost $r_i C$), plus the advertising income from sold out-links minus the advertising costs of bought in-links. The profit function (denoted here with $u$) is given by the following equation:

$$u_i \equiv r_i(c_i - C) + p_i d_i^{out} - \sum_{j \to i} p_j \qquad \textbf{(2)}$$

The Nash-equilibria of identity (2) represent a network of connected WGs characterized by two main features (assuming that WGs and Editors are a single entity):

a) WGs tend to buy links from other WGs with lower contents and
b) the higher the content of a WGs the more links it will buy from other WGs.

The result is also interesting, because it suggests that WG producers have a tendency to specialize in their business model. Certain WGs, the ones with low content specialize in selling links (i.e. traffic), while WGs with high content tend to buy links (advertise) in order to benefit from content sales. However, there are also WGs that do both, which is specific to the Web. This results in a network where the number of in-links correlates with the value of the corresponding WG.

### 6.4.2. Extensions of the basic model

The basic profit function (2) is extended to incorporate endogenous PPC and infinitely many WGs. The results arising from the discrete version of the model are valid for this infinite game. The model can be augmented to include reference links, Search engines and multiple content areas (i.e. Topics). Reference links are also created by Editors in order to increase the referring WGs' content with the assistance of the referred WGs (Mayzlin & Yoganarasimhan, 2008). In such case, apart from buying advertising in-links and selling advertising out-links, each WG is allowed to create a reference out-link to every other WG at maintenance cost $\kappa$. Reference links are part of the non-commercial Web can be symbolized as follows:

$$\boldsymbol{i \to Rj} \text{ represents a reference link for i to j and } \boldsymbol{j \to Ai} \text{ an advertising link from j to i.}$$

The motivation for and Editor to create reference links is to enrich her WG's content by referring to other WGs. The resulting generalized profit function accounts for the "effective" content term by two ways: (a) the WG's resident content $c_i$ and (b) the sum of the content of WGs linked to through reference links multiplied by a scaling constant $\beta \in [0,1)$. Hence, the total profit of WG $i$ is the following:

$$u_i \equiv f\left(d_i^{in_A}, d_i^{in_R}\right)\left(c_i + \beta \sum_{i \to Rj} c_j - C\right) - \kappa d_i^{out_R} + p_i d_i^{out_A} - \sum_{j \to Ai} p_j \qquad \textbf{(3)}$$

Katona and Sarvary (Katona & Sarvary, 2008), in order to solve this new form of the initial game, have incorporated in identity (3), a simplified flow of consumers in the network. They assume that $r_i = f\left(d_i^{in_A}, d_i^{in_R}\right)$ is the traffic (or the demand) for WG $i$. $f$ is a function of the WG's in-degrees and it is defined to be increasing and strictly concave in both advertising and reference links.

The last extension of the basic model (2) refers to the inclusion of Search Engines, allowing WGs to have multiple content areas, i.e. Topics. As Katona and Sarvary (Katona & Sarvary, 2008) explain:

*"Search engines (SE) play an important role in the formation of the network. If some consumers use SEs, then the number of visitors at a Web site does not only depend on the structure of the network but also on how search engines display the site in the result of a given search. Today's SEs use a twofold method to determine which pages and in what order to display the result of a search. On the one hand, they measure content directly, on the other hand, they measure content indirectly through the structure of the network, using methods such as Page Rank."*

The necessary generalization of the model focuses on letting the content of each WG to have up to D different dimensions or Topics $c_i^1, c_i^2, ..., c_i^D$ (e.g. news, music, business etc.). The proportion of consumers interested in different Topics is depicted by the weight vector $\mathbf{w}$. Hence, the consumer-specific content at WG $i$ is the scalar product $\mathbf{wc_i}$ and the weighted income for the Editor equals $r_i\mathbf{wc_i}$. The revised for multiple Topics total profit function (3), equals:

$$u_i \equiv r_i(\mathbf{wc_i} - C) + p_i d_i^{out_A} - \sum_{j \to i} p_j \text{ (4)}$$

where $p_i = \frac{\delta q_i r_i}{(d_i^{out}+1)}$ and $q_i = q(\mathbf{wc_i})$ is an increasing function of average content.

Profit function (4), still without the presence of Search Engines, results in the same equilibrium as the one described for (2), if content is replaced with the weighted average content.

Now it is assumed that $b$ proportion of consumers is navigating according to the random-surfer process described in (1) and the remaining $1 - b$ consumers use a Search Engine in every navigation step. More precisely, the Search Engine indexes the WGs with the $s$ highest content parameters in every dimension and directs consumers to one of these with probability proportional to their Page Rank. This mechanism is very much alike to the KKPS model (Kouroupas, Koutsoupias, Papadimitriou, & Sideri, 2005a). The vector $\mathbf{C_i}$ is defined to be the content vector $\mathbf{c_i}$, truncated by the Search Engine by setting to zero the dimensions that are not included in the top $s$ ranks. If the modified average content is $E_i \equiv b\mathbf{wc_i} + (1 - b) \mathbf{wC_i}$ then the total profit for WG $i$ is given by:

$$u_i \equiv r_i(\mathbf{E_i} - C) + p_i d_i^{out_A} - \sum_{j \to i} p_j \quad \text{(5)}$$

where $p_i = \frac{\delta q_i r_i}{(d_i^{out}+1)}$ and $q_i = q(.)$ is an increasing function of the modified average content $E_i$.

Profit function (5) and the resulting Nash-equilibria describe the most comprehensive version of the Katona-Sarvary model:

*"Clearly, with a single content area, the existence of a search engine does not matter qualitatively. It simply makes the "divide" between low and high content pages more pronounced... the equilibrium graph show that the sites with the highest $E_i$ will have the highest in-degree and Page Rank... Specifically, there is an incentive to specialize in a certain content area in order to be one of the top sites of a particular dimension and, in this way maximize the "specialization reward". On the other hand, this incentive to specialize decreases as the average content of a site is higher, since a high average content site does not have to allocate all its resources to one dimension, it can afford to diversify its content. Thus, we would expect sites with low total content to specialize, while those with high general content to diversify. However, as more and more people use search engines the advantage from high average content disappears and ultimately all sites compete for higher content in a specific area."*

### 6.4.3. Results

The main results of the Katona-Sarvary model can be summarized as follows:

- In all equilibria, both advertising and reference links direct to higher content WGs, verifying the practical importance of in-degree criterion as the basis of many search algorithms (e.g. Google).
- Contrastingly, the pattern of out-links is different for both advertising and reference links. WGs tend to purchase advertising links from lower content WGs. The higher content WGs prefer to create more reference out-links.
- In the presence of search engines, the above patterns become more pronounced.
- The degree distribution of in- and out-links is a scale-free power-law distribution with an

exponent of around 2, compatible to the empirical features of the Web structure.

Building on the Katona-Sarvary model, (Kominers, 2009) examines the strategic production of sticky content in commercial WGs that generate revenues from both selling services and selling links.

### 6.5. The Dellarocas-Katona-Rand model

Recently, the formation of hyperlinks, a fundamental characteristic of the Web, became the central of business controversies. As traditional content creators (e.g. newspapers and TV) are loosing a big part of their revenue streams from User-Generated substitutes (e.g. blogs and micro-blogs), Platforms, Search Engines and Reconstructors, are raising regulation issues in free reference linking. Characteristically, the Associated Press CEO, Tom Curley pointed out that[16]: *"Crowd-sourcing Web services such as Wikipedia, YouTube and Facebook have become preferred customer destinations for breaking news, displacing Web sites of traditional news publishers. We content creators must quickly and decisively act to take back control of our content."*

Oppositely, it is argued that Aggregators create exploitable traffic for content creators and all online content must be open on the Web with permanent links so it can receive in-links, since links are a key to efficiency on creating and finding information. Furthermore, the recipient of links is the party responsible for monetizing the audience it brings and there are business opportunities to add value atop the link layer (Jarvis, 2008). However, the economic implications of reference links on attention and revenue have not yet analyzed. It is an important challenge for Web's future to understand how the micro and macro structures of the "Link economy" influence consumer's utility, competition and social welfare in the economy as a whole.

Dellarocas, Katona and Rand (Dellarocas & Katona, 2010), in their ongoing work analyze, for the first time, the economic implications of free reference hyperlinks placement to content nodes. They argue that: *"This work offers a step in the direction of understanding the complex interplay between content and links in settings where a set of nodes compete for traffic and make strategic investments in both content and links to maximize their revenues. We are interested in answering both micro level questions of interest to firms (e.g., what is the optimal way in which content organizations can combine content creation and hyperlinking to optimize revenue? how can content organizations best leverage and respond to the presence of aggregators?) as well as macro level questions of interest to regulators (e.g., what are the implications of strategic hyperlinking for quality of content that is available to consumers? are aggregators beneficial or harmful to society?)."*

Contrastingly, to Web-based content network analysis, which employs empirical methods (for example: (Huberman & A. Adamic, 1999), (Huberman, Pirolli, Pitkow, & Lukose, 1998)) and makes no assumptions about the individual agents, Dellarocas et al (Dellarocas & Katona, 2010) explicitly model both Navigators and Editors as adaptive utility-maximizing entities. Based on the Katona-Sarvary model (Katona & Sarvary, 2008), they extend the existing literature of strategic network formation (for a review see (Jackson, 2008)) by accounting for simultaneous and interdependent node-level strategic decisions about both node properties and links.

The basic assumptions of the model are presented in the first Subsection 6.5.1. Subsection 6.5.2 briefly states the main results of the model.

### 6.5.1. Assumptions

Dellarocas et al (Dellarocas & Katona, 2010) extend and the Katona-Sarvary model in the media industries by revising Web traffic generation and the formation of Reference links. In particular:

    A. The simplistic assumption that the traffic is generated by the random-surfer model (Katona &

---

[16] http://mashable.com/2009/10/09/ap-news-corp-pay-us/%20

Sarvary, 2008) is replaced by a more realistic anchor-traffic provision. Dellarocas et al (Dellarocas & Katona, 2010) model Navigators as consumers, which want to maximize the utility they enjoy from online content while minimizing the undertaken cognitive cost. Navigators are used to select "anchor-WGs", a small number of WGs from which they start their news consumption (Purcell, Rainie, Mitchell, Rosenstiel, & Olmstead, 2010). Their traffic patterns depend also on outside alternative information sources (e.g. by watching TV) offering a direct connection of the Web to traditional economy.

B. The Katona-Sarvary model assumed that each WG is allowed to create a reference out-link to every other WG at maintenance cost $\kappa$. In the new specification of Dellarocas et al linking exhibits both benefits and costs for link sources and targets. For the link source, the main cost stems from the fact that placing a link to a WG of better content will make some Navigators to follow that link, leaving no revenue to the source WG. On the contrary, including a link to a better WG, Editors substitute their effort to create new content and their WGs become more attractive. The link target faces the opposite trade-off. The benefit of a reference link increases the possibility to attract more visitors, but decreases its relative attractiveness as an anchor node.

The Dellarocas-Katona-Rand model includes Navigators and Aggregators in general. Navigators maximize their utility for any bit of information per unit of attention and Aggregators maximize their pay-offs by placing links to the best available WGs. The analysis is focused on three major implications of Aggregators' function on the content ecosystems:

• Facilitate access to high quality online content. As result of this, increase the average attractiveness of the Web ecosystem compare to the outside alternative media (e.g. radio, newspapers) and the profits of the high quality WGs.
• Despite the fact that Aggregators do not produce original online content but only links, compete content WGs for advertising revenue. This fact decreases content WGs' pay-offs and weakens their incentive to produce quality content in the future.
• The formation of limited links or hierarchical presentations of them (e.g. Search Engine's results) by Aggregators, accelerate the competition among content WGs. As (Dellarocas & Katona, 2010) argue: *"This increases the quality of available content, which is good for consumers, but in most cases further decreases the profits of content sites because it forces them to overproduce. Interestingly, however, the additional competitive pressure that is brought forth by the presence of aggregators makes it more likely that sites will form link equilibria to alleviate the pressure."*

### 6.6. Results

The Dellarocas-Katona-Rand model is the first and important step towards understanding the positive and negative effects of "Link economy". The major conjecture of their on-going work is that: *"link equilibria often do not form, even though their formation can lead to higher aggregate profits and better content. This, in the view of the authors constitutes a negative side-effect of the culture of "free" links that currently pervades the web…"*

Briefly, their basic results are the following:
• Links among peer content producers can increase firm profits by reducing competition and duplicate effort.
• Links only form if competition among WGs is not too tough.
• Linking can sustain market entry of inefficient players.
• The main benefit of Aggregators to content producers comes from traffic expansion.
• The presence of Aggregators incurs social costs that must not be overlooked.
• Aggregators increase competition among content WGs.

## 7. Market regulation and antitrust issues

In the last scene of this "Link economy" play, Google lost substantial ground and reconsiders its strategy. Particularly, in May 2011, Google failed in the Court of Appeal in Brussels to reverse the court ruling that forced her in 2007 to remove links and snippets of articles from French- and German-language Belgian newspapers. The decision was justified on the fact that Google generated revenue without compensating publishers for providing their content[17]. At the same time, Google has shut down its Google News Archives digitization program, which aimed to make the world's newspaper archives accessible and searchable online. Released resources will be directed to a Web platform (e.g. Google One Pass) that enables publishers to sell content and subscriptions directly from their own sites. Despite this strategic set back, Google remains by far the dominant player in the Search Engine market (Pollock, 2010) and growing in a number of other businesses (e.g. Android). Actually, together with less than ten mammoth firms, in global scale, dominate the Web economy. Indicatively, Apple has more than a 70% share of paid music downloads in the European Union[18], eBay has more than a 90% share of auction site page views in France, Germany, Italy, Spain and the UK and Facebook has more than a 50% share in most countries for which there are data. These firms also have shares in antitrust markets that rival those held by Microsoft. These business practices in the Web raise antitrust concerns for competition issues and demand for regulation like in the past happened with global giants like Microsoft (Economides, 2001). Antitrust regulations are also likely under European Community law, which impose significant scrutiny on firms that have market shares as low as 40%. After 2007, together with competition issues, privacy worries brought in the spotlight as Users, nongovernmental organizations, academia (Lazer et al., 2009), government agencies and the media started focusing on the vast amount of personal behavior data that online advertising firms are storing and exploiting.

During last two years, came to the fore various important regulation issues about the Web. In January 2009, introduced in US Congress a comprehensive legislation[19] designed to address vulnerabilities to cyber crime, global cyber espionage, and cyber attacks. The Secretary of Commerce would be given access to any information *"without regard to any provision of law, regulation, rule, or policy restricting such access."* The bill would also give the President new authority to *"declare a cybersecurity emergency and order the limitation or shutdown of Internet traffic to and from any compromised Federal Government or United States critical infrastructure information system or network."* In August 2010, Google and Verizon unveiled a proposal[20] to maintain an open Internet while creating room for a broadband network of premium services. The proposal came as a response to public interest groups and lawmakers lobbying for the US government to mandate net neutrality. Finally, in December 2010, the Federal Communications Commission decided to create two classes of Internet access, one for fixed-line providers and the other for the wireless providers like Verizon[21]. This controversial "net semi-neutrality" ruling enables partial traffic discrimination in wireless Internet and initiated a new round of scientific and legal controversies.

In this Section, we examine the basic antitrust issues raised by the "information gatekeepers" of the Web (i.e. Search Engines) and the "infrastructure gatekeepers" of Internet (i.e. ISPs).

### 7.1. Antitrust issues in the Search Engine market: the Pollock model

Ruffus Pollock[22] is one of first researchers of the Web economy who is fully aware of the underlying technologies. In particular, he is an active programmer in the filed of Linked Data and related technologies and co-founder of the Open Knowledge Foundation[23]. His theoretical work (Pollock, 2010)

---

provides what is, to the best of our knowledge, the first formal analysis of the wider Search Engine market and its welfare implications. The main focus of analysis is to demonstrate that the already large levels of concentration in the Web search market are likely to continue; to identify the negative and positive consequences of this result and to propose a set of regulatory interventions that decision makers could employ to address these. According to Pollock (Pollock, 2010) the Search Engine (SE) market is characterized by two stylized facts: (a) a cost structure, which involves high fixed costs and low marginal costs and (b) pure quality competition for Users (i.e. zero prices and no User heterogeneity), that is likely to feature very high levels of concentration and under-provision of quality by a single dominant firm. He argues that since the market mechanism cannot provide socially optimal quality levels, there is space for regulatory engagement. Regulatory policies may involve the funding of basic R&D in Web search, or more drastic measures like the division of SEs into two separate parts: "software" and "service".

The first Subsection describes the characteristics of pricing and cost structures in the SE market. Subsection 7.1.2 states the basic assumptions of the model, while Subsections 7.1.3 and 7.1.4 analyze welfare and regulation issues, respectively.

### 7.1.1. Pricing and costs in the Search Engine market

Before constructing the economic model, Pollock analyzes the pricing strategies and costs of the SE market in the Web. First, the basic input for SEs (i.e. Web pages) is provided for "free". Second, SEs offer their service for "free" to all Users. It is crucial to understand the pricing policy of SEs. As Pollock (Pollock, 2010) explains:

- *"... the use-value of a search engine (the value of a query) is likely to be very heterogeneous (both across users and time) and hence may be difficult to price "well"".*
- *"... search engines are essentially (meta-)information providers supplying users with information about where other information is located. Hence, charging for their service (i.e. charging users for queries) would suffer from all the classic Arrovian difficulties, most prominently that the value of a given query is often highly uncertain before it is performed."*
- *"... charging users would necessitate significant transaction costs on two main counts. First, in relation to administration of charges (processing and payment). Second in maintaining an effective exclusion regime which prevented those who had not paid for use for gaining access, directly or indirectly, to the search engine's search results."*
- *"... search engines have an alternative method to direct charging for generating revenue from users: selling users' 'attention' (and intentions), generated via the use of the search facility, to advertisers."*

In today's complex Web, running a SE service requires high investments in capital and R&D. The costs to develop and maintain a competitive SE are primarily fixed. On the other hand, the marginal cost of serving one additional User (Navigator or Editor) is almost zero. This structure of high fixed costs in investment and direct supply together with almost zero marginal costs is similar to natural monopolies in utilities markets.

### 7.1.2.Assumptions

*(a) Users: Search Engine intermediates between Navigators and Editors*
The Pollock model categorizes Web Users to "Users", which navigate and search the Web (Navigators) and for Amateur Editors uses the term "content providers". Advertisers pay the SE to participate in the Web. Advertisers and the SE are the only Professional Editors in this "three-sided" setup. Navigators are looking for content, which is created by Amateur Editors. Amateur Editors, who want to attract more Navigators, are not buying traffic or in-links by the SE or other Editors like in the Katona-Sarvary model (Katona & Sarvary, 2008). There is no direct linkage or circular flows among all Web Users. The SE intermediates between Advertisers and Amateur Editors and between Advertisers and Navigators. Amateur Editors provide online content to index and receive traffic through the SE. Navigators receive search results without financial compensation, but by providing their attention, while Advertisers pay advertisement space in order to attract traffic (Figure 9).

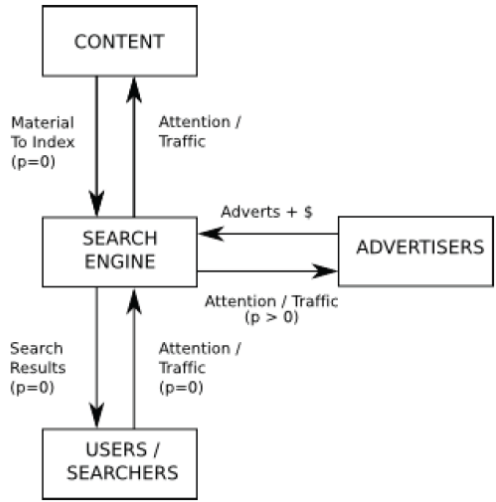

**Figure 9: The structure of the SE business (Pollock, 2010)**

The Pollock model divides Users in four categories: Navigators, Search Engines, Amateur Editors and Advertisers. Navigators-Queries are contracted in a single graph like in the KKPS model (Kouroupas, Koutsoupias, Papadimitriou, & Sideri, 2005a) and there are not specific assumptions for Topics and WGs. Despite the fact that the Pollock model involves a "three-sided" setup, there is only one financial flow between the SE and Advertisers (Figure 9). Hence, the commonly used in such cases "two-sided" markets analysis[24] is not appropriate. The Pollock model is build up on the standard toolkit of oligopolistic competition, particularly, on the models of Bertrand competition and vertical product differentiation (Shaked & Sutton, 1983), (Sutton, 1991). The basic assumptions of the Pollock model can be summarized as follows:

- Amateur Editors (called here "content providers", C[25]) are ignored as strategic agents because their content is available to all SEs for free.
- Search Engine (S) quality is reflected in a single variable $v$, which all Navigators value positively (increasing utility). Later in the model it is assumed that there is a single SE. SE do not charge Navigators. Its core costs are primarily fixed and the marginal cost of an additional User (Navigator, Amateur Editor and Advertiser) is approximately zero. The cost of supplying a given quality is independent of the number of Users.
- Consumer and producer surplus are accorded equal value in the social welfare function.

---

[24] For a detailed description refer to Section 2.6.
[25] The full set of symbols employed by the Pollock model is presented in Table 3.

**Table 3: The set of symbols employed by the Pollock model**

| | |
|---|---|
| $t$ | Navigators |
| $S$ | 1, …, N Search Engines |
| $v^i$ | The quality of a SE i |
| $p_u^i$ | The price that SE i charges Navigators |
| $q_u^i$ | $= q_u^i(v^i, v^{-i}, p_u^i, p_u^{-i})$ *and if* $p_u^i = 0$ *equals* $q_u^i(v^i, v^{-i})$ <br> The total user demand for search engine i |
| $\alpha$ | $= \alpha(t, v^i, q_u^i)$ *or simply* $\alpha(t, v, q_u)$ <br> the advertising revenue generated by Navigator t at SE i |
| $R_A$ | Total advertising revenue |
| $c$ | $= c(v, q_u)$, "core" or "User" costs |
| $c_A$ | $= c(v, q_u)$, "advertising" costs |
| $C$ | $= c + c_A$, total costs of a search engine |
| $\Pi_S$ | $= R_A - C$, Profits of a SE |
| $R_U$ | The total additional revenue accruing to advertisers from Navigators as a result of their advertising |
| $\Pi_A$ | $= R_U - R_A$, Total advertising profits |
| $U_A$ | Utility from advertising |
| $U_S$ | $= U_A - R_U$, Utility from search |
| $U_U$ | $= U_S + (U_A - R_U)$, Total utility of Navigators |

If the SE index *i* is dropped, the utility of a Navigator from using the SE is the following:

$$U_t = u(t, v, p_u) \qquad \qquad \textbf{(1)}$$

and by assuming that SE do not charge Navigators ($p_u = 0$) (1) becomes:

$$U_t = u(t, v) \qquad \qquad \textbf{(1)}'$$

The profits of a SE are given as the difference between revenues and cost:

$$\Pi_S = R_A - (c + c_A) \qquad \qquad \textbf{(2)}$$

and if $v$ is inverted and taken as a function of demand $[v = v(q_u)]$ and define $\bar{p}(q_u) = R_A(q_u)/q_u$, equation (2) can be written as:

$$\Pi_S = \bar{p}(q_u)q_u - C(q_u) \qquad \qquad \textbf{(2)}'$$

which is similar but not identical to the vertical product differentiation problem. Here, $\bar{p}$ is the implicit price charged to a Navigator through advertising. Navigators are not making choices based on prices but on quality. This set up is alike the classic Bertrand model with SEs competing on quality instead of price (Pollock, 2010). Assuming that total demand $q_u$ is a simple scalar equal to the measure of Navigators whose utility is positive and inverting $q_u(v)$ to get $v$ as a function of $q_u$, profit function (2) can be re-written as:

$$\Pi_S = R_A(v, q(v)) - c(v, q(v)) - c_A(v, q(v))$$
$$= R_A(v) - c(v) - c_A(v) \qquad \textbf{(2)}''$$

The profit maximization problem for a monopolistic SE is to select the quality level $v^M$ that maximizes $(2)''$. The optimal level of $v^M$ satisfies the following First Order Condition:

$$R'_A(v) = R_v + R_q q' = c' + c'_A \quad \textbf{(2.1)}''$$

where subscripts symbolizes partial derivatives and " $'$ " total derivatives.

The above near-monopoly model is a satisfactory approximation to reality in most markets of Web search. Pollock's results are robust to variations in the model structure given the underlying zero-Navigator price/quality competition, but this is not the case for strong contestability.

### 7.1.3. Welfare

The social welfare function $W$ is defined as:

$$W = Utility\ of\ Navigators + Profits\ of\ SE + Profits\ of\ Advertisers$$
$$= U_U + \Pi_S + \Pi_A = U_S + (U_A - R_U) + R_A - (c + c_A) + (R_U - R_A)$$
$$= U_S + U_A - (c + c_A) \quad \textbf{(3)}$$

and if advertising has a neutral effect on consumer's utility ($U_A = 0$) and $c_A = 0 \Leftrightarrow R_A = 0$, then equation (3) simplifies to:

$$W(C) = U_S(v, q(v)) - c(v) \quad \textbf{(3)}'$$

Then if equation $(3)'$ is maximized with respect to $v$, the social optimal level of quality of SE, $v^W$ is obtained from the following formula:

$$U'(v^W) = U_v + U_q q' = c' \quad \textbf{(3.1)}'$$

Pollock in order to answer how optimal is monopoly, compares the socially desirable level of search quality given by $(3.1)'$ and the monopolistic outcome in $(2.1)''$. He examines various cases of cost, revenue and quality patterns and concludes that:

*"Overall, it is likely, in our view that the under-provision effect dominates. First, the indirect effect of search engine quality on utility is likely to grow at least as fast, if not faster, than its effect on revenue. Second, the direct effect of quality on utility is likely to positive (and substantial) while the direct on revenue will be negative. Third, and least importantly, search engines have to bear advertising related costs, which increase their costs compared to the direct funding case and therefore reduce the quality provided".*

### 7.1.4. Regulation

The Pollock model described the reasons of increasing concentration in the Web search market. The under-provision of quality occurs primarily due to two basic reasons: the "social-private" gap ("*the benefits of an extra unit of search quality to society are less than those extracted in the form of advertising revenues*") and the "distortion" effect ("*quality acts as substitute for advertising*"). Thus, since the market mechanism cannot provide socially optimal quality levels, there is space for regulatory engagement. Pollock proposes the funding of basic R&D in Web search, which will provide scientific knowledge to all firms. In the same line, is the provision of computing grid and search index infrastructure upon which developers could experiment and build innovative algorithms. This regulatory proposal supports the provision of a personal grid workspace (g-work)[26] to (potentially) every physical and legal entity in order to access structured information and inter-create (Vafopoulos, Gravvanis, & Platis, 2006). The negative "distortion" effect could be anticipated by a set of rules and controls on the quality of SE's rankings and their connections to personal, social and business interests. A more drastic measure would be the division of the Web search service into two distinct parts: "software" and "service". As Pollock explains:

*"The "software" includes all the main software used to run the system, including the ranking algorithm. The "service" side involves all the infrastructure, data-centres, support systems etc, which run the software and actually respond to users' queries."*

## 7.2. Net Neutrality

Vint Cerf, one of the basic architects of the Internet and co-inventor of the Internet Protocol testified in Congress at 2006 that:

*"The Internet was designed to maximize user choice and innovation, which has led directly to an explosion in consumer benefits. The use of layered architecture, end-to-end design, and the ubiquitous Internet Protocol standard, together allow for the decentralized and open Internet that we have come to expect. This created an environment that did not require Tim Berners-Lee to seek permission from the network owners before unveiling a piece of software enabling the World Wide Web."* (Cerf, 2006).

Tim Berners-Lee, inventor of the Web and also a strong advocate of net neutrality, highlighted the importance of Internet infrastructure on the Web User experience (Berners-Lee, 2010).

As online access is becoming a universal right (Huffington Post, 2011), net neutrality is one of the main factors that determine the quality of this right. Thus, neutral access to online information is an issue of fundamental importance in the Web Science agenda. In this section we analyze the economic aspects of net neutrality including the effects in the social welfare. In the first subsection net neutrality is defined. Subsection 7.2.2 describes the opposition about net neutrality. The last subsection briefly discusses related economic modeling.

### 7.2.1. Definition

The provision of Internet services is considered to be *"neutral"* if Internet Users should pay ISP(s) only for the right to access the network at their end ("one-sided pricing"). In contrast, the access is characterized as *"non-neutral"* if Editors and developers of Internet applications should also pay ISP(s) for the "right" to reach Navigators and other Internet Users ("two-sided pricing"). The question is which of the two pricing policies maximize the social welfare.

### 7.2.2. The opposition

Providers of residential broadband access argue that in order to ensure efficient network use, including the high investment costs of upgrading, they ask for flexibility in managing and pricing Internet traffic. In particular, ISPs prefer to charge the providers of Internet content and applications based on the offered level of QoS (*"non-neutral"* access).

The Google–Verizon proposal (Google, 2010) attempted to change the focus of opposition from residential broadband access to the mobile phone-based connections. The proposal received many negative reactions. In the same line, Berners-Lee (Berners-Lee, 2010) argues that:

"*Many people in rural areas from Utah to Uganda have access to the Internet only via mobile phones; exempting wireless from net neutrality would leave these users open to discrimination of service. It is also bizarre to imagine that my fundamental right to access the information source of my choice should apply when I am on my WiFi-connected computer at home but not when I use my cell phone.*"

In the academia, the majority of scholars believe that ISPs, in the absence of network neutrality restrictions, will have high motives to discriminate or even exclude from their network some of the producers of Internet content and applications (for various point of views on the issue see (Schwartz & Weiser, 2009)). This discriminatory policy in access and pricing could be transformed to significant societal cost and serious threats in the amount of innovation and participation in the markets of Internet applications and WGs. According to Schwartz and Weiser (Schwartz & Weiser, 2009) the proponents of restrictions *"... fear that as incumbents expand the deployment and use of higher-priced "private networks" to handle all types of IP traffic (voice, video and data), they may under-invest in facilities for the public Internet in order to steer users to the higher-priced alternatives."*

---

[26] For a detailed description refer to Section 8.3.

### 7.2.3. A two-sided market analysis of Net Neutrality

A "non-neutral" Internet access raises six basic issues in the economic level (Economides & Tåg, 2007):

1. Two-sided pricing on the Internet.
2. High possibility for prioritization of information packets.
3. Identity-based discrimination in delivering information packets. In such case, ISPs can determine which of the firms in an industry sector on the other side of the network will get priority and therefore win.
4. New and innovative firms with small capitalization will actually be excluded from the prioritization auction.
5. ISPs have huge incentives to favor their own content and applications rather that those of independent firms.
6. "This can result in multiple fees charged for a single transmission and lead to a significant reduction in trade on the Internet".

Network neutrality has proven to be stimulating for innovation and inter-creativity in the application level. Of course, it is not coming without costs since it reduces network providers' incentives to innovate at the network level and to deploy network infrastructure (Economides, 2005). In this debate, Internet and Web scholars should evaluate which strategy minimizes risks and results larger total benefits for the society, despite the fact that scientific literature on net neutrality regulation is still on its infancy both in economic and legal studies. Further investigation is needed on a wide range of issues such as the impact of Net Neutrality regulation on innovation among content providers, non-linear platform pricing and congestion and broadband penetration (Economides & Tåg, 2007).

Recently, Economides, one of the protagonists in Network Economics research, analyzed the economic effects of Net Neutrality. In particular, Economides and Tag (Economides & Tåg, 2007) focused their study on the issue of one-sided versus two-sided pricing. They believe that this issue should play a larger role in the debate than the other issues (admittedly important) including the exclusion of content providers, QoS variations, dynamic investment incentives and price discrimination. Their model explicitly allows monopoly and duopoly access providers to charge a positive fee to content and applications producers and compared to the case of a regulator who chooses the fee to content providers to maximize the total surplus. (Economides & Tåg, 2007) demonstrate that under realistic parameter ranges, *"the regulator will choose a negative fee to content providers while a monopolist or duopolists will choose positive fees."* Also, *"...for some parameter values, society is better off in terms of total surplus at net neutrality rather than either the monopolist's or duopolists' choices of positive fees to content providers. However, there are also parameter ranges for which the opposite result is obtained."*

### 7.3. The Web in a crossroad

The growing influence of a small group of colossal Web firms in almost every aspect of the service sector steers innovation and productivity but raises concerns about the effect of their business practices in personal and total utility. Their excessive market power in search, commerce, auctions, social networking and other, seems to erect barriers of entry for new firms and to threat the Internet's and Web's building blocks and Users' privacy. Regulators are facing the options to divestiture the monopoly into separate firms, to unbundle WGs or to license proprietary interfaces to potentially competing platforms.

Apart from these repressive measures, there are public policies that support the creation of new knowledge and businesses in related fields. For instance, the funding of basic R&D in Web technology and standards combined with the provision of public or communal data, computing and networking infrastructure will enrich competition and innovation. Relatively, the potential of the Web as a development force, both in developed and developing economies, is examined in the next section.

## 8. Web-based development: brief overview and major challenges

In a plethora of studies, initially Internet and later the Web, were analyzed as the new communication and cooperation tools, which just extend existing capabilities. During the last decade, with the advent of the participatory Web, it was realized the change in paradigm that the new ecosystem brings. The Web has been evolved to a powerful mean of inter-creativity and drastically effects from a single business to a group of nations (e.g. recent revolutions in Arab nations[27]). Although, the Web as a stand-alone technological artifact is ethically and politically neutral, different power groups can use it in order to play a decisive role in anticipating or accelerating mass phenomena. At the core of this function is the fact that the Web facilitates the massive creation, editing and distribution of information in any digital format around the globe. Hence, it dramatically increases the significance of information as a factor of personal, social and economic change. Each nation or group of Users incorporates this new feasibility space in a different direction. For instance, the Chinese and Turkish governments increased their efforts to control online information flows. On the contrary, the Finnish government introduced broadband access as a legal right for all citizens[28] and the European Commission included in its research and policy agenda the online access to structured Public Sector Information[29].

The present section is not, by any means, a thorough review of the ICT- or Web-for-Development literature. Its unique goal is to highlight the new opportunities and challenges that the Web ecosystem contributes in the development process. Section 8.1 discusses the ICTs' role in relation to social inequality. Section 8.2 highlights the development drivers in the networked information economy. The next section analyzes a minimal framework for Web-based development policies. Section 8.4 briefly describes two representative Web-based policies in action and the last Section offers a concluding discussion.

### 8.1. ICTs' role in relation to social inequality

The commercialization of Information and Communication technologies (ICTs) during 1980s rejuvenated the discussion about the role of technology in economy and society. At the beginning, most of the scholars, in the absence of real and massive cases of study, chose to take one of the two sides: most of things will change for the good or for the bad. In that period, conspiratorial and catastrophological scenarios co-existed with hopes about eternal prosperity that the cyberspace will bring. One of the issues often discussed was the evaluation of ICTs' role in relation to social inequality (W. Chen & Wellman, 2005). Specifically, the equalization hypothesis specified that ICTs decrease social inequalities, while the opposite assertion is examined by the amplification hypothesis. The transformation hypothesis shifts the discussion to the practical stake of the Web ecosystem in the social welfare.

*(A) The Equalization hypothesis: ICTs decrease social inequalities*

According to optimists, ICTs will reduce social inequality because they have fundamental impact to economic growth, social development, cultural diversity and the empowerment of less favored regions and disadvantaged people (W. Chen & Wellman, 2005). In such hypothesis, ICTs technologies would offer the opportunity in poor countries to leapfrog into the online world by adopting the latest technologies of the developed world. Moreover, ICTs could lead to democracy and more effective government (see for instance (Negroponte, 1995) and (Barlow & Birkerts, 1995)).

---

Based on the equalization hypothesis, a long range of ICT-for-Development (or ICT4Dev) policies has evolved during the last decades. According to Heeks (Heeks, 2008, 2009) in the initial period (from mid-1950s to late-1990s) related policies were concentrated on computing for back-office applications in large government and private organizations in developing countries. The first main period (from late-1990s to late-2000s) of intense ICT4Dev investments was basically driven by the wide dialogue in academia and politics about the *digital divide* (see for instance (Norris, 2001), (DiMaggio & Hargittai, 2001) and (Van Dijk & Hacker, 2003)) and the Millennium Development Goals that have been set by the United Nations (Sachs, 2005). One of the typical policy implementations was the creation of tele-centres, which initially provided information on development issues such as health, education, and agricultural issues into poor communities and later Internet wired and wireless access combined with some governmental services (Heeks, 2008, 2009). The current second era or ICT4Dev 2.0, ranges from late-2000s and onwards and is characterized by the fact that mobile devices are replacing the tele-centre as the focus of functionality (for a review on recent studies of mobile phone use in the developing world see (Donner, 2008)). This era also includes the rapidly growing initiatives for providing online Open Government Data as a stimulus to more energetic involvement and collaboration between producers and consumers of Web Goods.

*(B) The Amplification hypothesis: ICTs increase social inequalities*
The pessimistic view claims that ICTs are reflecting existing or even amplifying economic and social disparities. This happens because only few disadvantaged people and communities can actually participate into the global networked ecosystem due to existent limitations both in physical infrastructure (e.g. telecommunications) and social institutions (e.g. education system) (W. Chen & Wellman, 2005). Additionally, the first-movers competitive advantage can be also effective in the case of developed countries, which are reinforcing political and economic inequalities in the online world (Castells, 2001). Particularly, the unequal access to ICTs resources may increase the social exclusion and the deficit in human capital concentration in few people and groups (Bucy, 2000).

*(C) The Transformation hypothesis: the Web ecosystem has transformative power*
Extensive research efforts showed that the Web ecosystem is a highly dynamic and complex system, which is affected and affects the economic and social discourse (see for instance (Berners-Lee et al., 2006)). As it is argued in this article, Web's impact mainly stems from its transformative power in the production, dissemination and consumption of knowledge. Considering the fact that knowledge is a primer factor in social welfare and justice, the black-or-white approach of the equalization and amplification hypotheses imposes a simplistic dilemma. The one-dimensional direct connection of ICTs with social inequalities should be now replaced by the more relevant question *"what changes need to be incorporated in the Web ecosystem to best serve humanity?"*

## 8.2. Development drivers in the networked information economy

*(A) The practical potential*
The first step in answering related questions should be the identification of fundamental connections between the Web functions and economic development. Relatively, Benkler (Benkler, 2007) has established a systematic approach on how the networked information economy changes the feasibility set for human welfare. He claims that for both liberal and social-democratic analytical frameworks
*"...the availability of information from nonmarket sources and the range of opportunities to act within a nonproprietary production environment improve distribution in both these frameworks, but in different ways."*
The central argument is that the Web depreciates the cost and the institutional barriers to increase the practical potential to exploit the inputs and outputs of the information economy (Benkler, 2007). On the other hand, the US- and EU-led international trade and intellectual property systems have been recognized as the main obstacles in the dissemination of benefits of the networked economy in global scale. Furthermore, Benkler (Benkler, 2007) highlights the strong dependence of the Human Development Index (HDI) (Anand & Sen, 1994) with the access of information, knowledge, and information-embedded goods (Figure 10). Sen's theory of economic development extends the traditional income-based metrics by anticipating deprivation not in terms of a lack of specific "endowments", but as the "unfreedom" to achieve certain "entitlements" (Sen, 1999). HDI has been proposed as an alternative to GDP development index, which includes quality of life measures like life expectancy at birth, adult literacy and school enrollment, and GDP per capita.

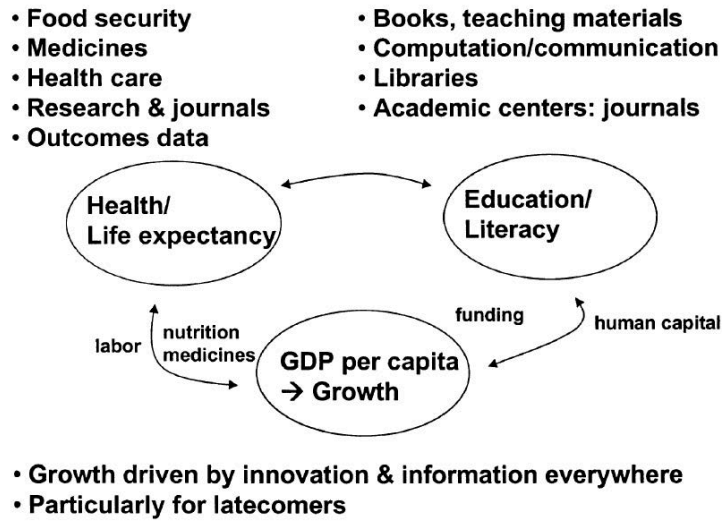

**Figure 10: HDI and information (Benkler, 2007)**

Benkler (Benkler, 2007) analyzes how the Web ecosystem facilitates positive influence to HDI by changing the feasibility space in creating and managing knowledge. In particular, he identifies the role of each type of main actors (i.e. government, NGOs, industry and individuals) in information and knowledge production across related sectors such as software, scientific publication, agricultural biotechnology, biomedicine and health.

*(B) Transparency and participation*
In the emergent Web ecosystem, Users have been facilitated with a greater participatory role in making the culture they occupy (Benkler, 2007). This culture is more transparent and open to any particular individual. Transparency and participation in sharing and creating knowledge could benefit the provision of life-critical functions (e.g. health, clean water and basic literacy) and human capital accumulation (e.g. education, innovation and creativity). Web-based development policies in developing countries are mainly focused on the first aspect, while developed countries are now concentrating their efforts to promote public data and services in the Web as basic infrastructure. Before going into a more detailed description of the undertaken actions, it is useful to present a minimal framework of development policies in the Web.

### 8.3. A minimal framework for Web-based development policies
The second step in understanding the Web's developmental potential is to consider a minimal framework of relevant policies. ICTs-based initiatives for development are focused on specific aspects of the digital divide such as broadband access, mobile services, technological literacy and knowledge sharing. But what these approaches are lacking of is an orchestrated provision of all tools as a personalized creative space. As Berners-Lee et al (Berners-Lee et al., 2006) explain that:

*"…while still preserving the relation between the invariants of the Web experience and the particular context of an individual's use that empower him or her to claim a **corner** of cyberspace and begin to use it as an extension of personal space. Given that, on the Web, the relevant information is likely to be highly distributed and dynamic, personalisation is expected to be one of the big gains of the Semantic Web, which is pre-eminently a structure that allows reasoning over multiple and distributed data sources."*

In this direction, (Vafopoulos, 2005) and (Vafopoulos et al., 2006) had initiated an analytical framework that describes an integrated bundle of access and processing infrastructure coupled with user-centric data and services in the Web for every physical and legal entity. The proposed *personal grid workspace (g-work)* is defined to have four interconnected aspects: (a) digital storage, (b) network traffic, (c) processing power and (d) one-stop data and services in the Web.

G-work is the public infrastructure part of this "corner of cyberspace" that will empower Users to exploit the benefits of the Web ecosystem. At the Web 2.0 era, this empowerment is more crucial than ever because the increasing complexity of online usage and data management weakens asymmetrically the power of Users compare to global firms and governments in the Web.

In the policy level, a series of governmental and non-governmental actions are partially providing one or more aspects but not the full g-work space. For instance, during 2009 the Finnish government guaranteed the network traffic for all citizens by making 1-megabit broadband access a legal right and the Greek Research and Technology Network is offering 50 GB of free digital storage to all members of the academic community.

In the business level, if g-work is considered as an alternative pillar in the provision of the four aforementioned aspects, could be part of the proposed antitrust policy against the growing market power of the gigantic Web firms. As it was mentioned in Section 3, most Reconstructors try to consolidate horizontally by adding more functionality and fragment vertically by providing services with multiple functionality like matchmaking, building communities and shared resources. An important part of their market power is built on the exploitation of personal data including demographics, navigational patterns and geo-location. Thus, one of the basic dimensions of the g-work is to enable Users, not only to create but also to *control* their personal data streams. This can be achieved by enforcing transparent exploitation mechanisms and organized exchanges of online information.

As Web's complexity grows in technological and organizational level, "mediators" dominate the information flows and weaken the influence of individual Navigators and Amateur Editors. Policies that provide an alternative public infrastructure to empower online individual Users will prevent excess market power and stimulate innovation in micro-scale. These policies have to be based on the basic principles of the Web (e.g. open standards, open source, open infrastructure and net neutrality) and if possible, to be orchestrated under a minimal common framework of provision.

### 8.4. Web-based policies in action

The final step in exploring Web-based development is to identify some representative initiatives. As the majority of scholars and decision makers have been convinced about the Web's catalytic power, the center of interest has moved in applying effective policies. According to recent reports, a small minority of ICT4D actions has proved to be sustainable over time (Walton, 2010). The key feature in this type of activities, as in the Web's massification, is the empowerment and the engagement of the beneficiary by the proper service provision. Indicatively, we describe two representative types of projects concerning Web-based development with different tasks. The *Web in Society* program was initiated by the World Wide Web Foundation to enable content sharing about life-critical functions through mobile phones in developing countries. In developed countries, the primary focus in content sharing is to unleash the economic potential of Open Government Data.

*(A) Web in Society*

In our point of view, one of the most promising projects, is the Web Foundation because it aims, by its inception, to coordinate the efforts of addressing three challenges for the Web, namely the Content, the Technology and the Research Gap. In particular, the Foundation's mission is concentrating in (a) making the Web better and more accessible, especially for the three-quarters of the world's population not yet using it, (b) empowering people in developing countries to build local capacity to accelerate development of valuable Web services and (c) providing global leadership on challenges associated with the Web.

The *Web in Society* action so far includes five initiatives[30]. First, the Web Index, which will measure its social, economic and political impact around the world in order to help governments, NGOs, and businesses to measure the effectiveness of their investments. Second, the Voice Browsing initiative will enable voice as a first-class interface to the Web, thus empowering billions of people who are not yet using the Web because they have only simple mobile phones, low literacy and/or disabilities. Third, the world-wide network of mobile Web incubation labs will stimulate mobile entrepreneurship by generating useful content and services for local communities. Fourth, the Web for Agriculture initiative will train and coordinate local developers to create and maintain Web-based platforms that help farmers, villages and extension agents in the African to share local innovations for growing vegetation in very harsh environments. Finally, the Open Government Linked Data initiative aims to improve transparency, accountability and competitiveness in global level and particular in Latin American, South and Southeast Asian and African regions.

*(B) Open Government Linked Data*

Recent economic analyses show that when re-usable information is provided to the public free cost, then individuals and private enterprises can take that information and create innovative and added-value products, which they can then market[31]. This economic activity stimulates the economy and also provides revenue to the government in the form of taxes. Much of this raw data could be integrated into new services such as car navigation systems, weather forecasts, or financial and insurance services. According to a survey conducted by the European Commission in 2006, the overall market size for Public Sector Information (PSI) in the EU is estimated at 27 billion euros (Dekkers, Polman, Te Velde, & de Vries, 2006). Europe first adopted the Directive 2003/98/EC on the re-use of PSI in 2003, the White House[32] noted that release of PSI serves *"to increase accountability, promote informed participation by the public, and create economic opportunity."* Data.gov.uk, the official Linked Data website of the UK government proclaims, *"Transparency is at the heart of this Government. Data.gov.uk is home to national & local data for free re-use"* and Europe, together with a growing number of governments and local authorities, are embracing similar strategy based on Linked Data technologies[33].

The economic literature related to Data-as-a-Service (e.g. Open Government Data, PSI) is still in its infancy. Recently, Pollock (Pollock, 2009b) argued that PSI is the basic infrastructure of the information economy, just as the supply of power, transport and communications networks are considered to be the prerequisites of development in the traditional economy. Hence, the maintenance, access and re-use of PSI are having a growing impact on the economy and society (Pollock, 2009b). Regarding the governance of PSI, it is proposed a transparent, independent and empowered regulatory framework. Particularly, *"... every public sector information holder there should be a single, clear, source of regulatory authority and responsibility, and this "regulator" should be largely independent of government."*

The charging scheme under investigation includes a combination of "updater" fees and direct government contributions with Users permitted free and open access in order to enjoy societal benefits from increased access to information-based services while imposing a limited funding burden upon government (Pollock, 2009b).

---

[30] Web Foundation just recently announced the project "Making the Web More Effective for Supporting Economic and Social Development" http://www.webfoundation.org/projects/web-study/
[31] An updated list of resources about the economic impact of open data is provided by the LinkedGov wiki http://wiki.linkedgov.org/index.php/The_economic_impact_of_open_data
[32] http://www.whitehouse.gov/sites/default/files/microsites/ogi-directive.pdf
[33] http://www.access-info.org/documents/documents/Beyond_Access_10_Aug_2010_consultationn.pdf

## 8.5. Discussion

Each major innovation – from typography to ipod and the Web– has spread unevenly throughout the globe. The varying ability of people to adopt and adapt new technology seems to be one of the main determinants of their prospects for development. At late-1990s, the resulting differences brought in the fore what is commonly referred to as the digital divide, but which is, in fact, a series of divides, disruptions and dynamics, presenting a complex mosaic of the penetration of technology.

Practical experience form development projects and extensive scientific investigation showed that the naïve binary approach of equalization or amplification of inequalities on account of ICTs should be replaced by more efficient conceptualizations.

The transformation hypothesis shifted the center of debate to more realistic questions about development such as *"what changes need to be incorporated in the Web ecosystem to best serve humanity?"*

Related to this approach, Benkler (Benkler, 2007) established a systematic approach on how the networked information economy updates the feasibility space for human welfare, by providing the practical potential to promote transparency and participation.

A minimal framework for Web-based development policies provides the second step in understanding the Web's developmental potential. The personal grid workspace (g-work) is defined by four interconnected aspects: (a) digital storage, (b) network traffic, (c) processing power and (d) one-stop data and services in the Web and is considered to be the public infrastructure part that will empower Users to exploit the benefits of the Web ecosystem.

In the level of realized policies, a series of governmental and non-governmental actions are partially providing one or more aspects but not the full g-work space. For example, the Web in Society program initiated by the Web Foundation to enable content sharing about life-critical functions through mobile phones in developing countries. In developed countries, the primary focus in content sharing is to unleash the economic potential of Open Government Linked Data. All these Web-based development policies have to be based on the basic principles of the Web (e.g. open standards, open source, open infrastructure and net neutrality) in order to optimize social benefits.

However, Peer production and Open Government Linked Data paradigms are not panacea for the economy and are not coming without limitations and costs. The basic limitations of these models arise from the fact that in the whole economy, geography still matters and in particular in the production of knowledge-intensive industries. Existing institutions such us intellectual property rights and trade agreements combined with oligopolistic market prices compose the second line of obstacles of Web-based development in global scale.

The main issue in Web-based development, as in the case of traditional development, is sustainability (Hopwood, Mellor, & O'Brien, 2005). That is, how the Web will become more useful and productive for the whole society with minimum negative externalities. But which could be the main sources of negative externalities in Web-based development? Is the Web polluting the environment and contributes to the climate change?

In our point of view, the most serious threat for a sustainable Web ecosystem is *"digital pollution"*, i.e. the abolishment of User's control and choice over her personal data online. The plethora of personal data spreading around the Web causes serious concerns about the existing and potential risks on personal and social level. In the extreme case, the ultimately polluted Web ecosystem is described as the one where all Users have been fully profiled. Proper "waste management" and "recycling policies" are needed to follow the development initiatives in the Web ecosystem.

Measuring the impact of Web-based development policies in economy is an open research issue. Further research also needed in adapting enforced policies to local conditions and social contexts.

## 9. Discussion and implications for further research

In the beginning of the present article we argue that the Web emerged as an antidote to the rapidly increasing quantity of accumulated knowledge and become successful because it facilitates massive participation and communication with minimum costs. Today, the enormous impact, scale and dynamism of the Web in time and space make very difficult (and sometimes impossible) to measure and anticipate the effects in human society. In addition to that, we demand from the Web to be fast, secure, reliable, all-inclusive and trustworthy. It is time for science to pay back the debt to the Web and provide an epistemological "antidote" to these issues.

Web-related studies can be found in many other disciplines than Computer science and Informatics. But the Web changes some of the underlying assumptions of the human society and that affect also it's engineering. On one hand, the Web offers more choices with less transaction costs in production and consumption. On the other hand, reinforces the abolishment of Users' control and choice over their personal data.

The complex interplay of social and technological entities occurring simultaneously and so massively in the micro and macro level is an uncharted ground of science. Web Science is a systematic effort to understand the Web ecosystem, model its stylized facts and engineer its future uses in more prosperous ways. One of the envelope questions of Web Science is "what changes need to be incorporated in the Web ecosystem to best serve humanity?" which can be further elaborated in two major research challenges for Web scholars:

1. to preserve and expand the fundamental right of equal and universal online access to information against restrictive political actions and oligopolistic business practices and
2. to accelerate socio-economic development by facilitating life-critical functions in the developing world and by enabling the publication, interlink and re-use of valuable datasets and services in the developed world.

The aforementioned research challenges imply that we need to better understand the updated value creation mechanisms, their relation to whole economy, their drawbacks and underlying costs. To be able to measure the impact of Web-based development policies in economy is an open research issue. Further research also needed in adapting enforced policies to local conditions and social contexts.

The Web as a dynamic evolutionary system share many things in common with live organisms. The interplay among function, structure and evolution, complexity, sustainability and biodiversity are just few out of many useful parallels for Web science research.

The unification and co-evolution of physical and online space challenges fundamental issues such as the notions of freedom, property, nation, innovation and "infrastructure gatekeepers". Traditional social values are changing content and ways of achieving them. New global human rights are emerging (e.g. online access).

Our new policies, practices and laws need to account by creation the transformative power of the Web ecosystem. To be able to preserve its variants and empower innovation, entrepreneurship and Users by providing them a sufficient level of public infrastructure and control over their personal data.

Instead of a concluding remark on Web's future, let me provide you an imaginary dialogue between Benkler and Wark.

Benkler (Benkler, 2007):
*"it (commons-based peer production) offers a new path, alongside those of the market and formal governmental investment in public welfare, for achieving definable and significant improvements in human development throughout the world."*
Wark (Wark, 2004):
*"It is not just information that must be free, but the knowledge of how to use it. The test of a free society is not the liberty to consume information, nor to produce it, nor even to implement its potential in private*

*world of one's choosing. The test of a free society is the liberty for the collective transformation of the world through abstractions freely chosen and freely actualised."*